\documentclass{article}

\usepackage[numbers]{natbib}
\usepackage{PRIMEarxiv}
\usepackage[T1]{fontenc}    
\usepackage{hyperref}       
\usepackage{url}            
\usepackage{booktabs}       
\usepackage{amsfonts}       
\usepackage{nicefrac}       
\usepackage{microtype}      
\usepackage{lipsum}
\usepackage{fancyhdr}       
\usepackage{graphicx}       
\graphicspath{{media/}}     
\usepackage[most]{tcolorbox} 
\usepackage{xspace}
\usepackage{color}
\usepackage{xcolor}
\usepackage{float}
\usepackage{multirow}
\usepackage{makecell}
\usepackage{amssymb}
\usepackage{pifont}
\usepackage{array} 

\usepackage{subfigure}
\usepackage{enumitem}

\usepackage{tikz}
\usepackage[edges]{forest}
\usetikzlibrary{external}
\definecolor{hidden-draw}{RGB}{20,68,106}
\definecolor{hidden-pink}{RGB}{255,245,247}

\newtcolorbox{remark}{
  colback=blue!5!white, 
  colframe=blue!75!black, 
  boxrule=0pt, 
  leftrule=2pt, 
  rightrule=2pt, 
  boxsep=5pt, 
  arc=0pt, 
  left=5pt, 
  right=5pt, 
  top=0pt, 
  bottom=0pt 
}

\newcommand{\mla}{Personal LLM Agent\xspace}
\newcommand{\mlas}{Personal LLM Agents\xspace}
\newcommand{\eg}{{\it e.g.,}\xspace}
\newcommand{\ie}{{\it i.e.,}\xspace}

\title{\mlas:\\Insights and Survey about the Capability, Efficiency and Security}  

\author{
\textbf{Yuanchun Li$^1$$^\dagger$, Hao Wen$^1$$^\ddagger$, Weijun Wang$^1$$^\ddagger$, Xiangyu Li$^1$$^\ddagger$, Yizhen Yuan$^1$$^\ddagger$, Guohong Liu$^1$$^\ddagger$,}\\
\textbf{Jiacheng Liu$^1$, Wenxing Xu$^1$, Xiang Wang$^1$, Yi Sun$^1$, Rui Kong$^1$, Yile Wang$^1$, Hanfei Geng$^1$,}\\
\textbf{Jian Luan$^2$, Xuefeng Jin$^3$, Zilong Ye$^4$, Guanjing Xiong$^5$, Fan Zhang$^6$, Xiang Li$^7$,}\\
\textbf{Mengwei Xu$^8$, Zhijun Li$^9$, Peng Li$^1$, Yang Liu$^1$, Ya-Qin Zhang$^1$, Yunxin Liu$^1$} \\ \\
$^1$ Institute for AI Industry Research (AIR), Tsinghua University \\
$^2$ Xiaomi AI Lab ~ $^3$ Huawei Technologies Co., Ltd. ~  $^4$ Shenzhen Heytap Technology Co., Ltd.\\
$^5$ vivo AI Lab ~ $^6$ Viomi Technology Co., Ltd. ~ $^7$ Li Auto Inc. ~ \\
$^8$ Beijing University of Posts and Telecommunications ~ $^9$ Soochow University\\ \\
$^\dagger$ Project Lead ~~~ $^\ddagger$ Section Lead\\
Contact: \texttt{liyuanchun@air.tsinghua.edu.cn} \\
{Website: \url{https://github.com/MobileLLM/Personal_LLM_Agents_Survey}} \\
}

\begin{document}
\maketitle

\begin{abstract}
Since the advent of personal computing devices, intelligent personal assistants (IPAs) have been one of the key technologies that researchers and engineers have focused on, aiming to help users efficiently obtain information and execute tasks, and provide users with more intelligent, convenient, and rich interaction experiences. With the development of the smartphone and Internet of Things, computing and sensing devices have become ubiquitous, greatly expanding the functional boundaries of IPAs. However, due to the lack of capabilities such as user intent understanding, task planning, tool using, and personal data management etc., existing IPAs still have limited practicality and scalability.

Recently, the emergence of foundation models, represented by large language models (LLMs), brings new opportunities for the development of IPAs. With the powerful semantic understanding and reasoning capabilities, LLM can enable intelligent agents to solve complex problems autonomously.
In this paper, we focus on \emph{\mlas}, which are LLM-based agents that are deeply integrated with personal data and personal devices and used for personal assistance.
We envision that \mlas will become a major software paradigm for end-users in the upcoming era. To realize this vision, we take the first step to discuss several important questions about \mlas, including their architecture, capability, efficiency and security.
We start by summarizing the key components and design choices in the architecture of \mlas, followed by an in-depth analysis of the opinions collected from domain experts.
Next, we discuss several key challenges to achieve intelligent, efficient and secure \mlas, followed by a comprehensive survey of representative solutions to address these challenges. 
\end{abstract}


\newpage

\tableofcontents
\newpage

\section{Introduction}



Science fiction has portrayed numerous striking characters of Intelligent Personal Assistants (IPAs), which are software agents that can augment individuals' abilities, complete complicated tasks, and even satisfy emotional needs. These intelligent agents represent most people's fantasies regarding artificial intelligence (AI). With the widespread adoption of personal devices (\eg smartphones, smart home equipment, electric vehicles, etc.) and the advancement of machine learning technology, this fantasy is gradually becoming the reality. Today, many mobile devices embeds IPA software, such as Siri \cite{siri}, Google Assistant \cite{google_assist}, Alexa \cite{alexa}, etc. These intelligent agents are deeply entwined with users, capable of accessing user data and sensors, controlling various personal devices, and accessing personalized services associated with private accounts.

However, today's intelligent personal assistants still suffer from the limitations of flexibility and scalability. Their level of intelligence is far from adequate, particularly evident in their understanding of user intent, reasoning, and task execution. 
Most of today's intelligent personal assistants are limited to performing tasks within a restricted domain (\eg simple functions in built-in apps). Once a user requests for tasks beyond these boundaries, the agent fails to comprehend and execute the actions accurately. Altering this circumstance necessitates a significant expansion of the agent's capability to support a broader and more flexible scope of tasks. 
However, it is difficult for current IPA products to support tasks at scale. Most of the today's IPAs require to follow specific predefined rules to complete tasks, such as developer-defined or user-demonstrated steps. Therefore, developers or users must explicitly specify which functions they wish to support, in addition to defining the triggers and steps for task execution. This approach inherently restricts the scalability to wider range of tasks, since supporting more tasks demands extensive time and labor cost. Some approaches have attempted to automatically learn to support tasks through supervised learning or reinforcement learning \cite{seq2act,glider,Liu2018ReinforcementLO}. However, these methods also rely on a substantial amount of manual demonstrations and/or the definition of reward functions.

The emergence of Large Language Models (LLMs) \cite{zhao2023survey} in recent years has brought brand new opportunities for the development of IPAs, demonstrating the potential to address the scalability issues of intelligent personal assistants. In comparison to traditional methods, large language models such as ChatGPT, Claude, and others have exhibited unique capabilities such as instruction following, commonsense reasoning, and zero-shot generalization. These abilities have been achieved through unsupervised learning on massive corpora (exceeding 1.4 trillion words) and subsequently fine-tuned with human feedback. Leveraging these capabilities, researchers have successfully adopted large language models to empower autonomous agents (aka. LLM agents), which aims to solve complex problems by automatically making plans and using tools such as search engines, code interpreters, and third-party APIs.

\begin{figure}[ht]
  \centering
  \includegraphics[width=16cm]{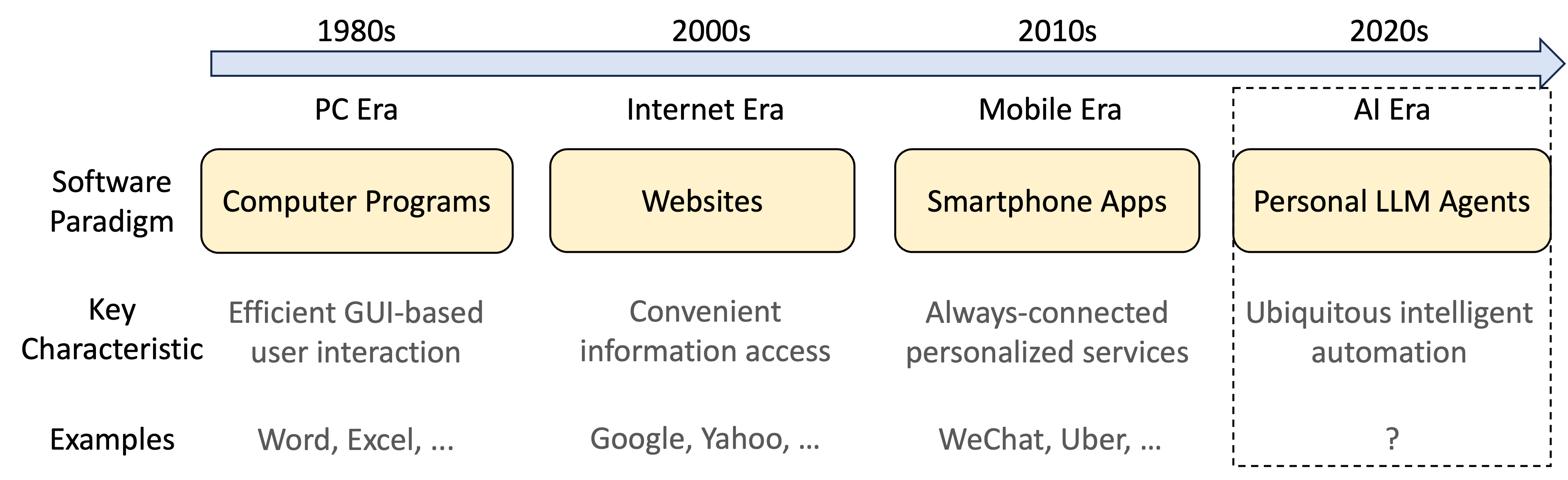}
  \caption{We envision {\bf \mlas} to become the dominating software paradigm for individual users in the upcoming era.}
  \label{fig:paradigm}
\end{figure}

As a unique type of intelligent agents,
IPAs also have the potential to be revolutionized by LLMs with significantly enhanced scalability, capability, and usefulness. We call such LLM-powered intelligent personal assistants as \textbf{\mlas}. As compared with normal LLM agents, \mlas are more deeply engaged with personal data and mobile devices, and are more explicitly designed for assisting people rather than replacing people.
Specifically, the primary way to assist users is by reducing repetitive, tedious, and low-value labor in their daily routine, letting the users focus on more interesting and valuable things, thereby enhancing the efficiency and quality of their work and life. 
\mlas can be built upon existing software stacks (\eg mobile apps, websites, etc.), while bringing refreshing user experience with ubiquitous intelligent automation abilities. Therefore, we expect \mlas to become a major software paradigm for personal computing devices in the AI era, as shown in Figure~\ref{fig:paradigm}.

Despite the promising future of \mlas, related research is still in its nascent stage, presenting numerous intricacies and challenges. This paper takes the first step to discuss the route map, design choices, main challenges and possible solutions in implementing \mlas.
Specifically, we focus primarily on the aspects related to ``\emph{personal}'' parts within \mlas, encompassing the analysis and utilization of users' personal data, the use of personal resources, deployment on personal devices, and the provision of personalized services. The straightforward integration of the general language capabilities of LLMs into IPAs is not within the scope of this paper.

We started by taking a survey with domain experts of \mlas. We invited 25 chief architects, managing directors, and/or senior engineers/researchers from leading companies who are working on IPAs and/or LLMs on personal devices.
We asked the experts' opinions about the opportunities and challenges of integrating LLMs in their consumer-facing products.
Based on our understanding and analyses of experts' insights, we summarized a simple and generic architecture of \mlas, in which the intelligent management and utilization of personal data (user context, environment status, activity history, personalities, etc.) and personal resources (mobile apps, sensors, smart-home devices, etc.) play the most vital role.
The ability to manage and utilize these personal objects differentiates the intelligence of \mlas. Inspired by the L1-L5 intelligence levels of autonomous driving, we also give an taxonomy of five intelligent levels of \mlas.

Our findings also highlight several major technical challenges to implement such \mlas, which can be categorized into three aspects including the fundamental capabilities, efficiency, and security \& privacy.
We further dive deeper into these aspects with detailed explanations of the challenges and comprehensive survey of possible solutions.
Specifically, for each technical aspect, we briefly explain its relevance and importance to personal LLM agents, then break it down to several main research problems.
For example, the foundamental capabilities for personal LLM agents include task execution, context sensing, and memorization. 
The efficiency of agents is primarily determined by the LLM inference efficiency, customization efficiency, and memory retrieval efficiency.
The security and privacy concerns of personal LLM agents can be categorized as data confidentiality, decision reliability, and system integrity.
For each research problem, we summarize the main techniques involved with the problem, followed by a brief introduction of the related work.
Due to the wide scope of the techniques in personal LLM agents, we only include the most relevant or recent works, rather than attempting to cover all related approaches.

The main content and contributions of this paper can be summarized as follows:

\begin{enumerate}
  \item We summarize the status quo of existing intelligent personal assistants in both industry and academia, while analyzing their primary limitations and future trends in the LLM era.
  \item We collect insights from senior domain experts in the area of LLM and personal agents, proposing a generic system architecture and a definition of intelligence levels for personal LLM agents.
  \item We review the literature on three important technical aspects of personal LLM agents, including foundamental capabilities, efficiency, and security \& privacy. 
\end{enumerate}

\section{A Brief History of Intelligent Personal Assistants}

\begin{figure}[ht]
  \centering
  \includegraphics[width=15cm]{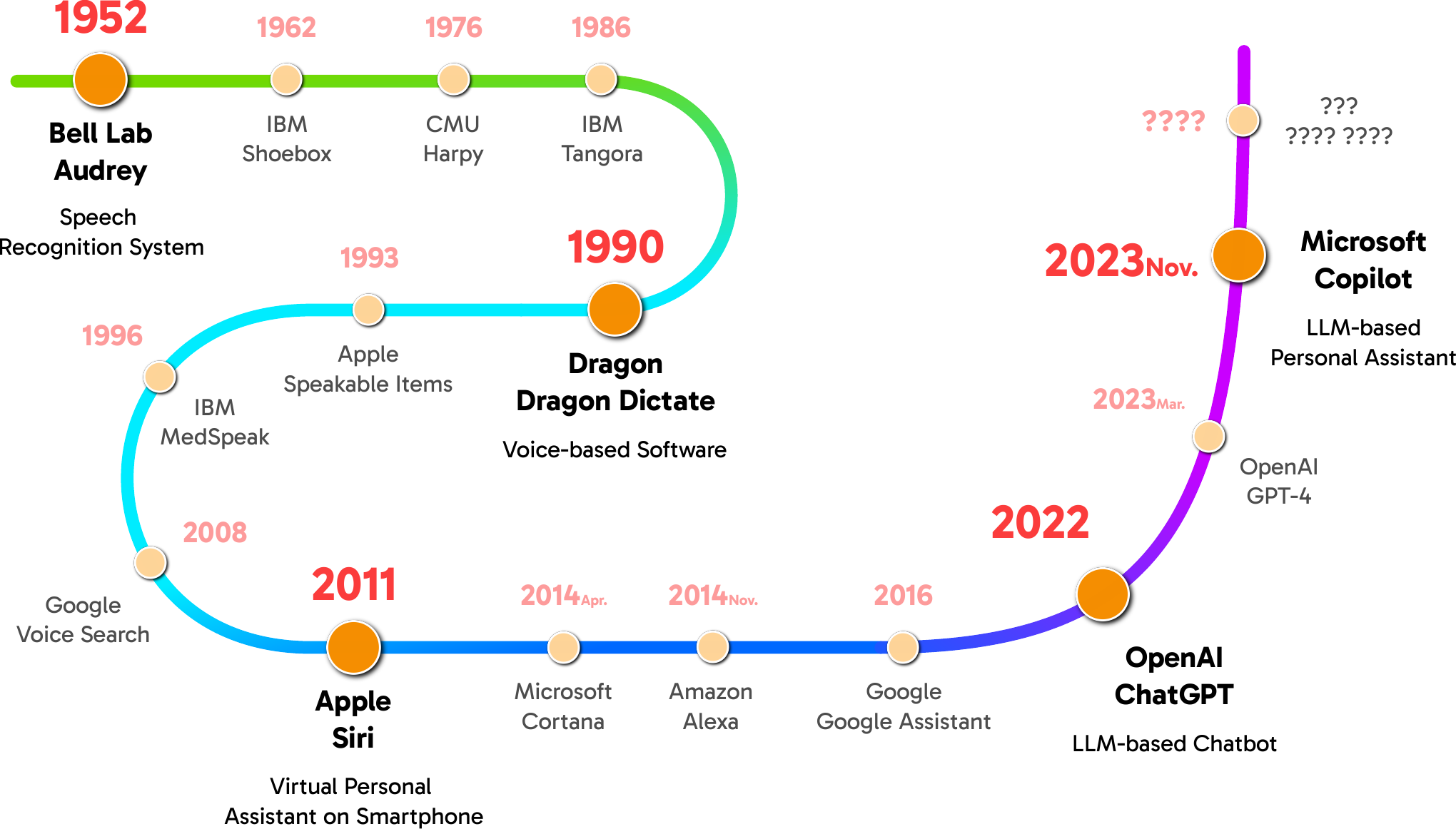}
  \caption{Major milestones in the history of intelligent personal assistants (IPAs). We mark different development stages with different colors, and some \textbf{significant or ground-breaking} events are highlighted with \textbf{bold text}.}
  \label{fig:timeline}
\end{figure}

\subsection{Timeline View of the Intelligent Personal Assistants History}

Intelligent Personal Assistants (IPAs) have a long history of development. We depict the rough timeline of the IPA history in Figure~\ref{fig:timeline}.
The development progress can be divided into four stages, each marked with a unique color in the figure.

\textbf{The 1st stage} spans from the 1950s to the late 1980s, which is mainly about the development of speech recognition techniques.
The early stage of speech recognition started from basic digits and words. Bell Laboratories developed ``Audrey'', which could recognize numbers 0-9 with about 90\% accuracy.
In 1962, the ``shoebox'' \cite{shoebox} system came out from Advanced Systems Development Division Laboratory at IBM, which was capable to recognize for up to 16 words.
From 1971 to 1976, the Speech Understanding Research (SUR) project, funded by the US Department of Defense, significantly advanced speech recognition technology.
The Harpy system \cite{lowerre1976harpy} was particularly representative, as it could understand sentences composed of 1011 words, equivalent to the proficiency of a three-year-old child.
In 1986, IBM developed the Tangora speech recognition typing system \cite{cerfdanon91_eurospeech}, capable of recognizing 20,000 words and offering predictive and error-correction capabilities. The Tangora system utilized Hidden Markov Models \cite{hmm}, requiring individual speaker training for voice recognition, with pauses between each word.

\textbf{The 2nd stage} covers the period from the 1990s to the late 2000s, since speech recognition started to be integrated into software for certain advanced functions.
In 1990, the ``Dragon Dictate'' software \cite{Bamberg1990TheDC} was released, which was the first speech recognition product for consumers. It was originally designed to work on Microsoft Windows, supporting discrete speech recognition.
``Speakable items'' \cite{Speakableitems} was introduced by Apple in 1993, enabling users to control their computer with natural speaking.
In 1996, IBM launched ``MedSpeak'' \cite{MedSpeak} for radiologists, which is also the first commercial product supporting continuous speech recognition.
Microsoft integrated speech recognition into Office applications in 2002 \cite{WinHEC2002},
and Google added voice search to Google Mobile App on iPhone in 2008 \cite{googleiphone}.

\textbf{The 3rd stage} extends from the early 2010s. In this period, always-on virtual assistant services began to appear on mobile devices such as smartphones and personal computers.
Siri \cite{siri}, widely considered as the first intelligent personal assistant installed on modern smartphones, was integrated into Apple's iPhone 4S in 2011. Since its launch, Siri has remained a key built-in software for Apple devices, including iPhones, iPad, Apple Watch, HomePod and Mac, continuously undergoing updates and iterations to incorporate new features.
Similar to Siri, many other virtual intelligent assistant started to appear in the period.
In 2014, Microsoft released Cortana \cite{cortana}, and gradually integrated it into desktop computers and other platforms.
Amazon released Alexa \cite{alexa} in the same year, which could complete tasks such as voice interaction, music playing, setting alarms, etc.
Beyond voice search, Google Assistant \cite{google_assist} was unveiled in 2016, supporting users to interact with both speaking and keyboard input.

\textbf{The 4th stage} started recently when LLMs start to draw attention from all over the world. Based on LLMs, there emerged many intelligent chatbots (\eg ChatGPT \cite{ChatGPT}), as well as some LLM-powered IPA software installed on personal devices (\eg Copilot \cite{Copilot}). The details of this stage will be covered in Section~\ref{sec: foundation_models}.

\subsection{Technical View of the Intelligent Personal Assistants History}

Since there are many aspects that can reflect the intelligence of personal assistants, we select one of the most important ability of Intelligent Personal Assistants, namely the task automation ability (following instructions and completing tasks), to be mainly focused on.
In the following subsections, we will introduce four main types of techniques to enable intelligent task automation in IPA. Note that these types of solutions have been developing concurrently, and there is no strict chronological order between them.

\subsubsection{Template-based Programming}

\begin{figure}[ht]
  \centering
  \includegraphics[width=12cm]{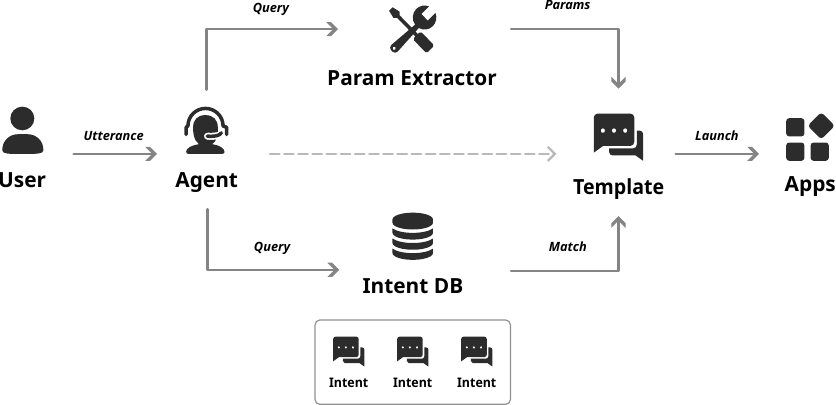}
  \caption{The workflow of template-based task automation.}
  \label{fig:temp_based}
\end{figure}

Most of the commercial IPA products support task automation through template-based approaches.
In these approaches, the functions that can be automated are predefined as templates, each of which usually contains the task description, related actions, example queries to match, supported parameters to fullfil, etc. Given a user command, the agent first map the command to the most relevant template, then follow the predefined steps to complete the task.
The workflow is illustrated in Figure~\ref{fig:temp_based}.

When using this method to automate tasks, app developers are required to follow the document of certain APIs (\eg the Google Assistant API \cite{google_assist}, SiriKit \cite{sirikit}, etc.) to create the template for each function they want to automate.
Besides, some approaches are proposed to enable end-users to create their own templates of tasks, such as the ``Shortcuts'' \cite{shortcuts} feature on iPhone devices, enabling the automation of repetitive operation sequences. Similar functions are also implemented in many products and academic research for the Android system, such as Tasker \cite{tasker}, Anywhere \cite{anywhere}, Epidosite \cite{li2017programming_iot} and Microsoft's uLink
\cite{ulink} system, etc.

The advantages of such template-based task automation method lie in its reliability and accuracy, since the steps in the template are deterministic and carefully programmed. However, its scalability is pretty limited, because of the relatively complex mechanism for supporting new tasks.
As a result, most apps, including the popular apps from large companies, do not support any automated task or only support some elementary ones, leading to very unflexible user experience. End-users can easilly give up the idea to use IPAs after several unsuccessful attempts \cite{ipa_experience, ipa_mixmethods, ipa_userexpect, alexa_siri_cortana}. This limitation poses a major obstacle to the further development of template-based intelligent personal assistants.

\subsubsection{Supervised Learning Methods}

To address the constraints of template-based IPA methods, researchers are actively investigating automated approaches for enhanced UI understanding and automation. Supervised learning offers a direct method for task automation by training models that predicts subsequent actions and states based on task inputs and current states. The main research questions include how to learn a representation of software GUI and how to train the interaction model.

The idea of learning an interaction model from human interaction traces is introduced in Humanoid \cite{li2019humanoid}, which aims to generate human-like test inputs based on the GUI layout information. Seq2act \cite{seq2act} firstly focused on the mobile UI task automation domain, where the natural language instructions need to be mapped to a sequence of actions that can be directly executed. The framework decomposed the problem into an \textit{action phrase-extraction} part and a \textit{grounding} part, both using the Transformer \cite{transformer} network.
Inspired by the success of pretraining in NLP, ActionBert \cite{He2020ActionBertLU} uses self-supervised pretraining to enhance the model's understanding of UIs. Specifically, to capture the semantics information of the UI switching actions, the model is designed to take a pair of UIs as input, and output embeddings of both UIs and individual components.
\citet{Fu2021UnderstandingMG} extended the concept of \textit{Words/Sentences} from NLP to \textit{Pixel-Words/Screen-Sentences}. By pre-training with visual atomic components (Pixel-Words), the PW2SS framework (Sentence Transformer) could accomplish various downstream GUI understanding tasks.
Aimed at better compatibility with the restricted resource on mobile devices, Versatile UI Transformer (VUT) \cite{Li2021VUTVU} was proposed to learn different UI grounding tasks within a single small model. It handles images, structures, and text-based types of data, using 3 task heads to support performing 5 distinct tasks simultaneously, including UI object detection, natural language command grounding, widget captioning, screen summarization and UI tappability prediction.
Based on the self-aligned characteristics between components of different modalities, UIBert \cite{uibert} presented a well-designed joint image-text model to utilize the correspondence, learning contextual UI embeddings from unlabeled data.
To address the problem of lacking UI metadata, such as DOM tree and view hierarchy, SpotLight \cite{Li2022SpotlightMU} introduced a vision-only approach for mobile UI understanding by taking screenshots and a region of interest (the ``focus'') as input. Composed of a vision encoder and a language decoder, it can complete tasks according to the provided screenshot and prompt.
Besides, Lexi \cite{Banerjee2023LexiSL} was proposed to leverage text-based instruction manuals and user guides to curate a multimodal dataset. By fusing text and visual features as input to the co-attention transformer layers, the model is pre-trained to form connections between text-based instructions and UI screenshots.
UINav \cite{Li2023UINavAM} utilized a referee model to evaluate the performance of the agent, immediately inform the users of the feedback. It also adopted demonstration augmentation to increase the data diversity.

As compared with template-based methods, supervised learning approaches have the potential to generalize to unseen tasks after sufficient training.
However, training the model typically requires a lot of high-quality human-annotated data. Given the diversity of tasks and apps in the real world, obtaining the training data that covers diverse use cases is challenging.

\subsubsection{Reinforcement Learning Methods}

Unlike supervised learning-based task automation approaches that require a large amount of training samples, reinforcement learning (RL)-based approaches allows the agent to acquire the capability of task automation by continuously interacting with the target interfaces. During the interaction, the agent gets feedback of rewards that indicate the progress of task completion, and it gradually learns how to automate the tasks by maximizing the reward payoff.

To train RL-based task automation agents, a reward function that indicates the progress towards task completion is required. World of Bits (WoB) \cite{wob} was proposed as a general platform for agents to complete tasks on the Web using keyboard and mouse. The platform came with a benchmark called ``MiniWoB'', containing tasks on a set of self-created toy websites with predefined rewards.
Glider \cite{glider} defines the reward function for real-world websites based on the semantic similarity between the task description and the UI action sequence, as well as the \textit{locality} and \textit{directionality} of the action sequence.

Another challenge of RL-based task automation is the huge action space and the sparse reward. A typical GUI-grounded task usually involves $5$-$10$ steps, each of which contains $10$-$100$ candidate actions, leading to a search space size of $10^5$-$100^{10}$. The task is completed only if the correct sequence of actions is taken. In order to tackle such challenge, many frameworks have been proposed.
\citet{Liu2018ReinforcementLO} introduced the method to use high-level ``\textit{workflows}'' to constrain the allowable actions at each time step. The workflows can prune out bad exploration directions, accelerating the agent's ability to discover rewards.
\citet{Gur2018LearningTN} \textit{decomposed} the complicated instruction into multiple smaller ones, and schedule a curriculum for the agents to gradually manage to follow an increasing number of sub-instructions. Besides, a meta-learning framework is also proposed to generate instruction-following tasks.
\citet{Jia2019DOMQNETGR} framed the actions of agent on the web into three distince categories, namely, DOM selection, token selection, and mode selection. What's more, a factorized Q-value function is designed, assuming the \textit{independence} of DOM selection and token selection.
Glider \cite{glider} achieves its goal of reducing action space with a hierachical policy, which contains a master policy to handle the overall navigation and sub-policies to deal with specific widgets.
\citet{rl_use_computer} proposed the framework to directly use mouse and keyboard to complete tasks instead of depending on the specialized action spaces, which simplifies the use of behavioural priors informed by actual human-computer interactions.

Similar to supervised learning methods, the RL-based methods also suffer from poor generalization ability. To achieve flexible and robust task automation, the RL agent needs to train on a large amount of tasks, each requires a well-designed reward function. Defining the reward functions for massive diverse tasks can be difficult.

\subsubsection{Early Adoption of Foundation Models}
\label{sec: foundation_models}

In recent years, pretrained large fundation models, represented by large language models (LLMs), have seen rapid development and brought new opportunities for personal assistants.

The scaling law \cite{kaplan2020scaling} for language models reveals the importance of increasing model parameters for improving model performance, followed by a bunch of models with billions of parameters.
The LLMs are typically trained with large-scale open-domain text data in an unsupervised manner, followed by instruction fine-tuning \cite{ouyang2022training} and reinforcement learning with human feedback (RLHF) \cite{christiano2023deep,ouyang2022training} to improve performance and alignment.
ChatGPT \cite{ChatGPT} unveiled by OpenAI at the end of 2022 is a milestone of LLM that demonstrated astounding question-answering capabilities.
By feeding simple task descriptions into the LLM as input prompts, the tasks and responses of LLMs can be easily customized. Besides, these models have also demonstrated robust generalization abilities across various language understanding and reasoning tasks.
ChatGPT itself can be viewed as an intelligent personal assistant that assist users by returning information in text responses.

Inspired by the capabilities of LLMs, researchers have attempted to let LLMs use tools \cite{schick2023toolformer} autonomously to accomplish complex tasks. For instance, such as controlling browsers \cite{nakano2022webgpt, Furuta2023MultimodalWN} for information retrieval and summarization, invoking robot programming interfaces for robot behavior control \cite{singh2023progprompt,zhen2023robot,huang2022language}, and calling code interpreters for complex data processing \cite{shen2023hugginggpt, Wang2023MathCoderSC,Rozire2023CodeLO, zhou2023solving}, among others. It is a natural idea to integrate these capabilities into intelligent personal assistants, enabling more intelligent ways to manipulate personal data, personal devices and personalized services.

There are already some commercial products that have attempted to integrate LLM with IPA. For instance, Microsoft's Copilot system \cite{Copilot} has integrated the capabilities of GPT-4 \cite{openai2023gpt4}, assisting users of Windows in automatically drafting documents, creating presentations, summarizing emails, and thereby enhancing user work efficiency. New Bing \cite{NewBing} also improves the experience of surfing the internet, providing a powerful efficient search engine which better understands what users want.
Similarly, Google has integrated LLMs (Bard \cite{bard}, Gemini \cite{gemini}) into the search engine to enable more convenient web search experience.
Smartphone companies including Huawei, Xiaomi, Oppo, Vivo have also integrated large models (PanGu \cite{PanguHuawei}, MiLM \cite{milm}, etc.) into their on-device IPA products. It is worth noting that some of them adopt solutions based on locally-deployed lightweight LLMs.
So far, most of these commercial products are just simple integration of the chat interfaces of LLMs into the personal assistants. Research about deeper functional integration will be discussed in Section~\ref{sec: task_auto}.

Despite exhibiting vast potential, this research direction is currently in an early exploration stage. There is still a substantial distance away from the ultimate goal of truly understanding and assisting users with intelligent agents. What's more, many issues related to efficiency, security and privacy have not been adequately addressed yet. The subsequent parts of this paper will systematically summarize and discuss the key issues in this direction.

\section{\mlas: Definition \& Insights}


Witnessing the great potential of LLM-based intelligent personal assistants and wide interests in both academia and industry, we take the first step to systematically discuss the opportunities, challenges and techniques related to this direction.

We define \textbf{\mlas} as a special type of LLM-based agent that is deeply integrated with personal data, personal devices, and personal services.
The main purpose of personal LLM agents is to assist end-users, helping them to reduce repetitive and cumbersome work and focus more on interesting and important affairs.
Following this definition, the generic automation methods (prompting, planning, self-reflection, etc.) are similar to normal LLM-based agents. We focus on the aspects that are related to the ``personal'' parts, such as the management of personal data, the use of smartphone apps, deployment to resource-constrained personal devices, etc.

We envision that \mlas will become a major software paradigm for personal devices in the LLM era. 
However, the software stack and ecosystem of \mlas are still at a very early stage. Many important questions related to the system design and implementation are unclear yet.

Therefore, we attempted to address some of the questions based on insights collected from domain experts. 
Specifically, we invited 25 experts who are chief architects, managing directors, or senior engineers/researchers from 8 leading companies that are working on IPA-related products, including smartphone personal assistants, smart-home solutions, and intelligent cockpit systems. 
We talked with them casually on the topics of \mlas and asked them several common questions, ranging from the application scenarios to the deployment challenges. 
Based on our discussion and collected answers, we summarize the insights into three subsections, including the key components of \mlas, a taxonomy of intelligence levels, and expert opinions about common problems.






\subsection{Key Components}

\begin{figure}[ht]
  \centering
  \includegraphics[width=11cm]{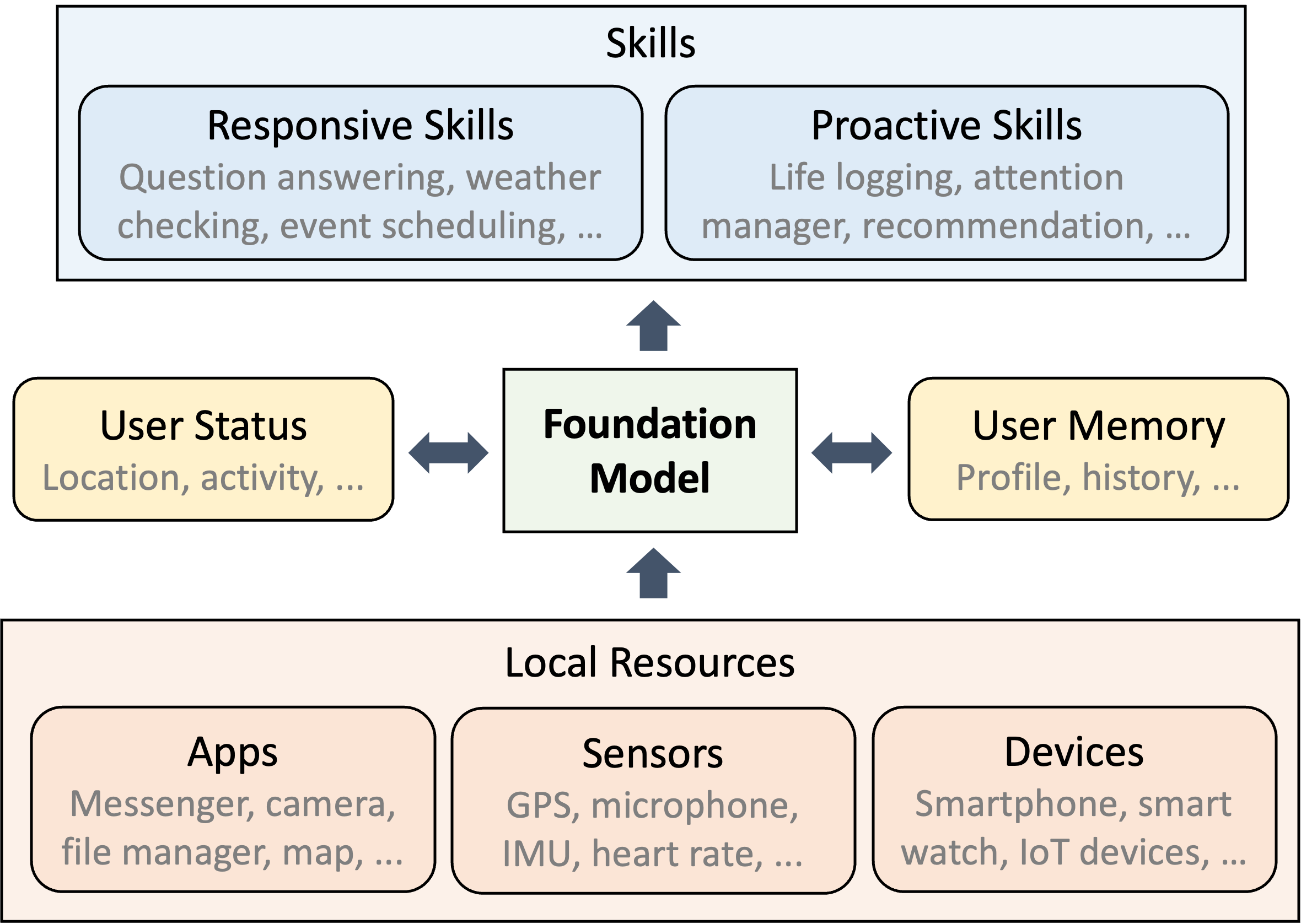}
  \caption{Main components of Personal LLM Agents.}
  \label{figs:architecture}
\end{figure}

Based on our discussions about the desired features of \mlas, we first summarize the main components to support such features, as shown in Figure~\ref{figs:architecture}.

Undoubtedly, the core of \mlas is a foundation model (large language model or other variants, we call it LLM for simplicity), which connects all other components.
Firstly, the LLM is the basis to support different skills for serving the users, including responsive skills that directly execute tasks as users requested (such as question answering, weather checking, event scheduling, etc.) and proactive skills that offer services without explicit user commands (such as life logging, managing user attention, activity recommendation, etc.).

Secondly, to support these skills, the LLM manages various local resources, including mobile applications, sensors, and IoT devices. For example, the agent may complete weather checking by interacting with a smartphone weather app.
Meanwhile, many people have mentioned the importance of \mlas to provide personalized and context-aware services. Therefore, the LLM should maintain the information about the user, including the current user context (status, activity, location, etc.) and historic user memory (profile, logs, personality, etc.). 
To manipulate these resources, contexts and memories, it is also desired to use dedicated management systems like vector databases in combination with the LLM.

The combination of these key components is analogous to an operating system~\cite{bokhari1995linux}, wherein:

\begin{enumerate}

\item The foundation model is like the kernel in traditional operating systems.
It is employed for systematic management and scheduling of various resources, thereby facilitating the functions of the agents.

\item The local resource layer is similar to the driver programs in traditional operating systems. 
In traditional OS, each driver manages a specialized set of hardware. While in \mlas, each local resource component manages a type of tool and provides APIs for the LLM to use.

\item User context and user memory correspond to the program contexts and system logs maintained during system operations.
These components form the basis for the agent to support personalized services.

\item The skills at the top layer are analogous to the software applications in traditional OS.
Similar to the installation and removal of applications, the skills of agents should also be allowed to be flexibly enabled or disabled.
\end{enumerate}



\subsection{Intelligence Levels of Personal LLM Agents}

\begin{figure}[b]
  \centering
  \includegraphics[width=12cm]{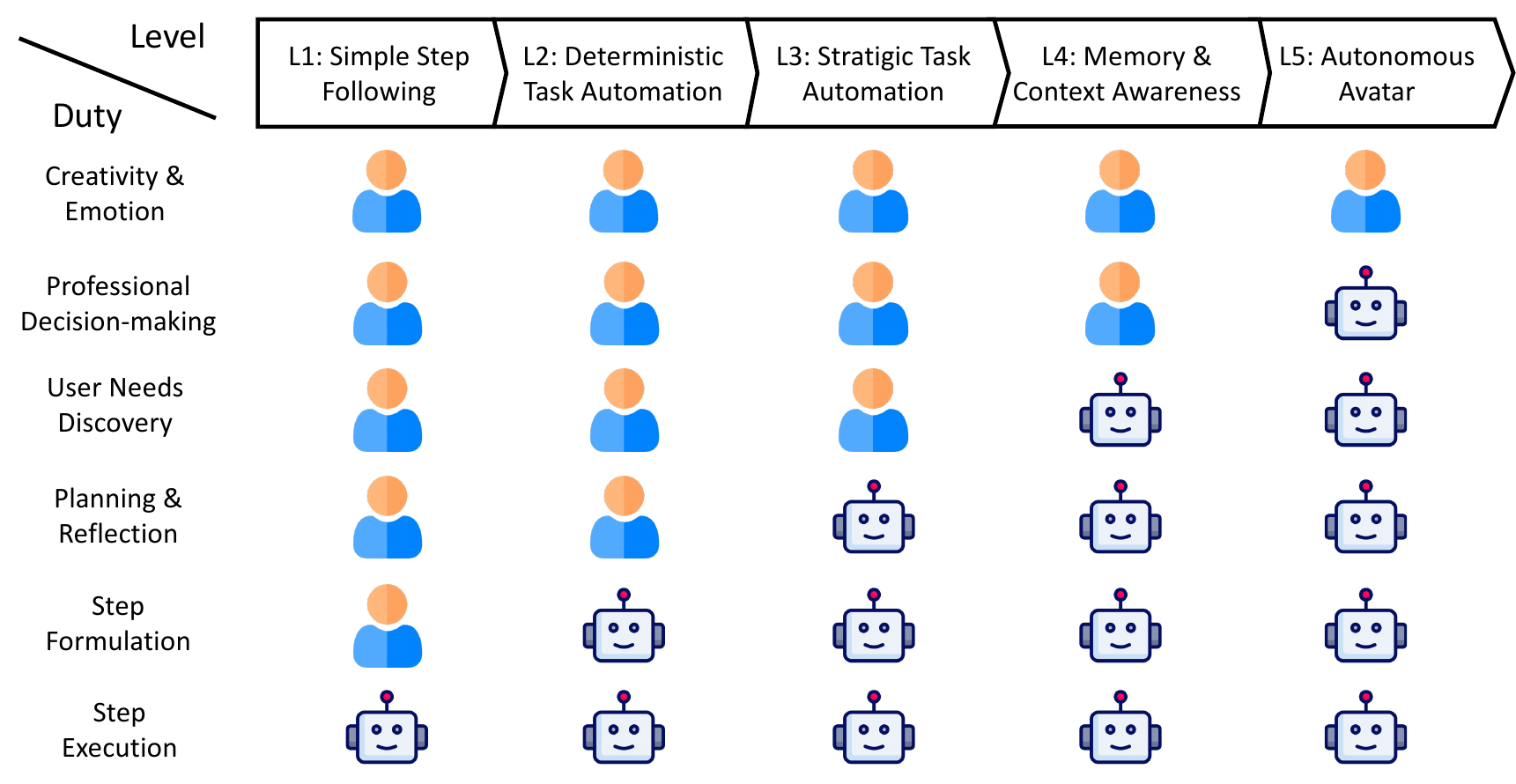}
  \caption{The duties of \mlas at different intelligence levels.}
  \label{figs:DutyLevel}
\end{figure}

The desired features of \mlas require different kinds of capabilities.
Inspired by the six levels of autonomous driving, we categorize the intelligence levels of \mlas into five levels, denoted as L1 to L5, as shown in Figure~\ref{figs:DutyLevel}. The key characteristics and representative use cases of each level are listed in Table~\ref{tab:intelligence_levels}.

\begin{table}[t]\small
  \centering
\caption{Different levels of intelligence for \mlas.}
  \label{tab:intelligence_levels}
  \begin{tabular}{m{1.7cm}|m{3.4cm}|m{10.2cm}}
  \toprule
  \textbf{Level} & \textbf{Key Characteristics} & \textbf{Representative Use Cases} \\ \midrule
  L1 - Simple Step Following & Agent completes tasks by following \emph{exact steps} predefined by the users or the developers.
  & - User: ``Open Messenger''; Agent opens the app named Messenger.\newline - User: ``Open the first unread email in my mailbox and read its content''; Agent follows the command step by step.\newline - User: ``Call Alice''; Agent matches a developer-defined template, finds Alice's phone number in the address book, and calls the number. \\ \hline
  L2 - Deterministic Task Automation & Based on the user's description of a deterministic task, agent \emph{auto-completes} the necessary steps in a predefined action space. & - User: ``Check the weather in Beijing today''; Agent automatically calls the weather API with parameter ``Beijing'' and parses info. from the response.\newline - User: ``Make a video call to Alice''; Agent automatically opens the address book, finds Alice's contact, and clicks on ``video chat''.\newline - User: ``Tell the robot vacuum to clean the room tonight''; Agent opens the robot vacuum app, clicks `schedule', and sets the time to tonight. \\ \hline
  L3 - Strategic task Automation & Based on user-specified tasks, agents \emph{autonomously plan} the execution steps using various resources and tools, and \emph{iterates} the plan based on intermediate feedback until completion. & - User: ``Tell Alice about my schedule for tomorrow''; Agent gathers tomorrow's schedule information from the user's calendar and chat history, then summarizes and sends them to Alice via Messenger.\newline - User: ``Find out which city is suitable for travel recently''; Agent lists several cities suitable for travel, checks the weather in each city, summarizes the information, and returns recommendations.\newline - User: ``Record my sleep quality tonight''; Agent checks every 10 minutes during sleep time if the user is using the phone, moving, or snoring (based on smartphone sensors and microphone), summarizes the information, and generates a report. \\ \hline
  L4 - Memory and Context Awareness & Agent senses user context, understands user memory, and proactively provides \emph{personalized} services at appropriate times. & - Agent recommends suitable financial products automatically based on User's recent income and expenses, considering User's personality and risk preference.\newline - Agent estimates User's recent anxiety level based on the conversations and behaviors, recommends movies/music to help relax and notifies user's friends or doctors depending on the severity.\newline - When a user falls in the bathroom, the Agent detects the event and decides whether to ask the user, notify the user's family members, or call for help based on the user's age and physical conditions. \\ \hline
  L5 - Autonomous Avatar & Agent \emph{fully represents} the user in completing complex affairs, can interact on behalf of user with other users or agents, ensuring \emph{safety} and \emph{reliability}. & - Agent automatically reads emails and messages on behalf of User, replies to questions without user intervention, and summarizes them into an abstract.\newline - Agent attends the work discussion meeting on behalf of the user, expresses opinions based on user's work log, listens to suggestions, and writes the minutes.\newline - Agent records User's daily diet and activities, privately researches or ask experts on any anomalies, and makes health improvement suggestions. \\ \bottomrule
  \end{tabular}
\end{table}

At each level, the user and agent are responsible for different duties.
At Level 1 (Simple Step Following), agents only take charge of step execution, and the other duties are in charge of the user.
For example, when users give the command, agents follow explicit steps defined by the developer or given by the user to complete the task.
The L1 agents do not have any ability of sensing or planning. Most template-based IPA products belong to this category.

As the intelligence level increases, the agents gradually take on more duties.
At level 2, the supported tasks are still deterministic (\ie involving a fixed sequence of actions to complete), but the detailed steps to execute each task are no longer given explicitly. The agents have to auto-complete the necessary steps based on the user's task description.
For instance, given a user query ``\texttt{How is the weather of Beijing today}'', the agent calls the weather API with \texttt{Beijing}'' as a parameter and retrieves weather information from the response.
Unlike the deterministic tasks at level 2, agents at level 3 can complete more complicated tasks that require strategic planning and self-reflection.
For instance, the command ``\texttt{Tell Alice about my schedule for tomorrow}'' needs the agent to determine how to gather the schedule information (\eg using the user's calendar and chat history) and how to inform Alice about the information (\eg summarizing the calendar events and sending via the messenger app). 
In these tasks, agents autonomously and iteratively generate and perform the execution plan based on intermediate feedback until completing the tasks.

The agents in L1-L3 work passively driven by the users' commands, while agents at level 4 can understand users' historical data, sense the current situation, and proactively offer personalized services at appropriate times.

With ultra intelligence at level 5, agents play the role of an Autonomous Avatar that can fully represent the user in completing complex affairs, thus users only need to focus on creativity and emotion.
Agents not only sense the current status, but also predict the users' future activities and take actions to facilitate them.
Beyond directly serving users, an Autonomous Avatar can also collaborate with other agents to alleviate the burden of their users' communication.
Moreover, the level-5 agents should be able to continuously improve themselves through self-evolution.


\subsection{Opinions on Common Problems}

Next, we report the aggregrated results of the experts' opinions towards several common questions.
The questions include the design choices and the potential challenges to deploy \mlas, as summarized in Table~\ref{tab:questions}.

We analyze the answers to the questions and summarize the following main takeaways.

\begin{table}[ht]
  \centering
  \caption{The common questions that we asked the domain experts. In Questions 1 to 6, we gave several common options for the experts to select/prioritize, while the experts were also allowed to give free-form answers. In Questions 7 and 8, the experts were asked to answer with text.}
  \label{tab:questions}
  \begin{tabular}{cp{14cm}}
  \toprule
  \textbf{ID} & \textbf{Question} \\ \midrule
1 & If the LLM is applied to personal intelligent agents, do you think it should be deployed locally or remotely? \\ 

2 & How do you think customized models tailored for different users or organizations should be implemented? \\ 

3 & For the LLM deployed on personal devices, which modality(ies) do you think needs to be supported? \\ 

4 & What do you think is the most important capability of LLMs for personal LLM agents? \\ 

5 & Considering the industry you are in, which ways of interaction do you think are the most promising for personal LLM agents? \\ 

6 & In the future development of personal LLM agents, which aspect is the most crucial? \\ 

7 & What features do you hope a future personal LLM agent can provide for you or your customers? \\ 

8 & When integrating LLM with personal devices, what challenges do you think will be faced? What are the most urgent technical issues that needs to be addressed? \\ 
  \bottomrule
  \end{tabular}
\end{table}

\begin{figure}[ht]
  \begin{minipage}[t]{0.49\linewidth}
  \centering
  \includegraphics[width=1.\linewidth]{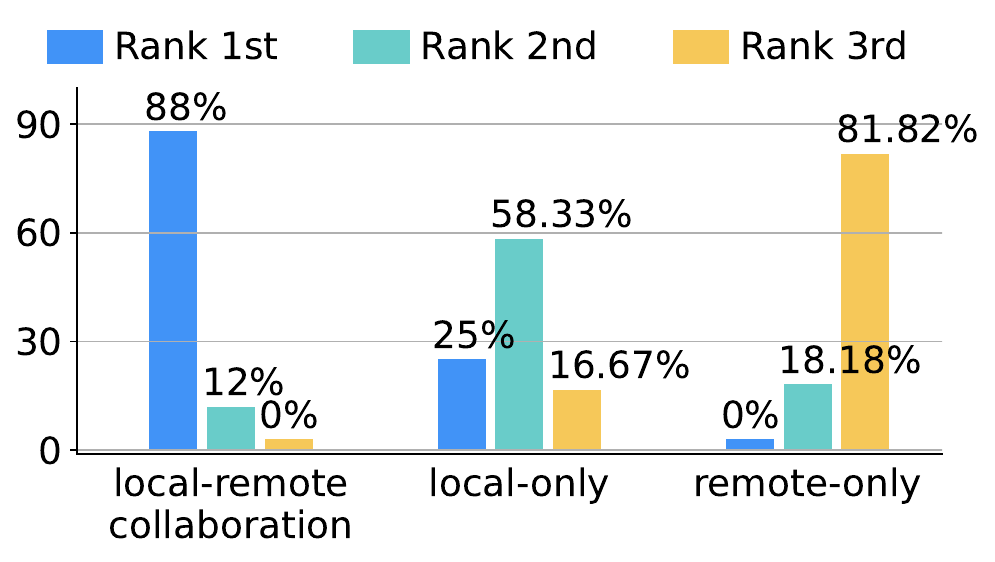}
  \caption{The vote distribution of different LLM deployment strategies in \mlas.}
  \label{figs:Q1}
  \end{minipage}
  \hfill
  \begin{minipage}[t]{0.49\linewidth}
  \centering
  \includegraphics[width=1.\linewidth]{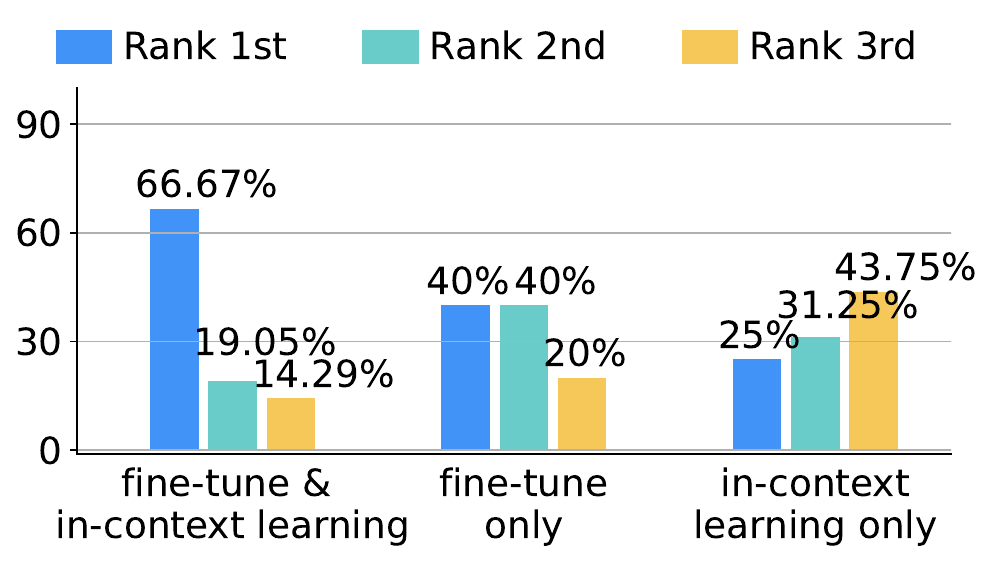}
  \caption{The vote distribution of different model customization methods for \mlas.}
  \label{figs:Q2} 
  \end{minipage}
\end{figure}

\textbf{Opinion 1 (where to deploy the LLM):}
\emph{Edge-cloud (local-remote) collaborated deployment of LLM is preferred, while existing cloud-only (remote-only) (\eg ChatGPT) is not a widely acceptable solution.}
As shown in Figure~\ref{figs:Q1}, 88\% of participants prefer an edge-cloud collaborated architecture, 58.33\% of them support local deployment, and 81.82\% of them are not satisfied with the existing cloud-only solutions.
Their main concerns are 1) the high latency of remote LLM service, 2) the privacy issue of transmitting personal data to the cloud, and 3) the huge cost of cloud-based LLM services.

\textbf{Opinion 2 (how to customize the agents):}
\emph{Combining fine-tuning and in-context learning is the most acceptable way to achieve customization.}
In \mlas, customizing the agent for different users and scenarios is considered necessary.
Figure~\ref{figs:Q2} shows that 66.67\% of participants support combining the advantages of both fine-tuning and in-context learning to reach personalization (L4 intelligence).
43.75\% of them do not believe L4 can be achieved by in-context learning; one possible reason is our participants are from the industry, thus they are more focused on the LLM for specific vertical domains where in-context learning hasn't received much attention. 

In questions 3-5, we ask participants to rank the options and the following tables (Table \ref{tab:Q3}-\ref{tab:Q5}) summarize their ranks.
Rank 1st-4th denotes the rankness of these options voted by the participants; for example, 72\% in Table \ref{tab:Q3} means that 72\% participants rank Text as their first preferred modality.
The ``score'' in each table is calculated based on the Borda Count~\cite{BordaCount}, where each candidate receives points equal to the average of the number of candidates they outrank in each ballot, with the lowest-ranked getting $2$ and the highest $n+1$ points, where n is the total number of candidates.
For instance, $4.56$ in Table \ref{tab:Q3} equals to $5\times72\%+4\times20\%+3\times0+2\times8\%$.

\textbf{Opinion 3 (what modalities to use):} 
\emph{The multi-modal LLM, especially Textual and Visual modalities, is desired for \mlas.} 
In our statistical result, Text is the most preferred modality just as the most popular LLMs used (\eg GPT series and LLaMA series).
The second-ranked Image option and the Video modality which is specifically mentioned by 20\% of the participants show that the visual modality plays a promising role in the future of personal LLM agents.

\begin{table}[ht]
  \centering
  \caption{The favored modalities to be used in \mlas.}
    \label{tab:Q3}
  \begin{tabular}{c|ccccc}
  \toprule
  \textbf{Options} & \textbf{Scores} & \textbf{Rank 1st} & \textbf{Rank 2nd} & \textbf{Rank 3rd} & \textbf{Rank 4th}\\ \midrule
  Text & 4.56 & 72\% & 20\% & 0\% & 8\%\\ \midrule
  Image & 3.64 & 4\% & 64\% & 24\% & 4\%\\ \midrule
  Voice & 3.18 & 16\% & 4\% & 60\% & 20\%\\ \midrule
  Sensors & 2.18 & 9.52\% & 14.29\% & 9.52\% & 66.67\%\\
\bottomrule
  \end{tabular}
\end{table}

\textbf{Opinion 4 (which LLM ability is the most crucial for IPA products):} 
\emph{Language understanding is considered the most important capability of LLMs, whereas the ability to handle long contexts is regarded as the most unimportant one.}
On the contrary, in academia, the capability to handle long context is regarded as very important and is extensively studied.
This different opinion originates from the specific vertical-domain LLMs our participants supposed and the general-purpose LLMs of academic researchers.
In vertical-domain LLMs, the queries and tasks from users are not very diverse, hence the capacity of long context is not that critical. 

\begin{table}[ht]
    \centering
    \caption{The importance ranking of LLM abilities for IPA products.}
      \label{tab:Q4}
    \begin{tabular}{c|ccccc}
    \toprule
    \textbf{Options} & \textbf{Scores} & \textbf{Rank 1st} & \textbf{Rank 2nd} & \textbf{Rank 3rd} & \textbf{Rank 4th}\\ \midrule
    Language understanding & 4.52 & 83.33\% & 8.33\% & 4.17\% & 4.17\%\\ \midrule
    In-context learning & 3.16 & 4.55\% & 50\% & 45.45\% & 0\%\\ \midrule
    Common sense reasoning & 3 & 8.33\% & 33.33\% & 29.17\% & 20.83\%\\ \midrule
    Long context & 1.8 & 5.56\% & 11.11\% & 16.67\% & 61.11\%\\
  \bottomrule
    \end{tabular}
\end{table}

\textbf{Opinion 5 (how to interact with the agents):} 
\emph{Voice-based interaction is the most popular way.}
Unsurprisingly, just like the existing virtual assistant Siri, mimicking the human communication method -- voice interaction is the most common and efficient choice.
Text-based chatbots and GUI rank second and third since most of the participating experts focus on mobile devices, \eg smartphones.
Virtual reality only obtains a $1.52$ score which is the lowest across all questions; this may stem from the high price of VR devices and the unsatisfied user experience of current VR techniques.

\begin{table}[ht]
  \centering
  \caption{The favored interaction method of \mlas.}
    \label{tab:Q5}
  \begin{tabular}{c|ccccc}
  \toprule
  \textbf{Options} & \textbf{Scores} & \textbf{Rank 1st} & \textbf{Rank 2nd} & \textbf{Rank 3rd} & \textbf{Rank 4th}\\ \midrule
  Voice interaction & 4.04 & 60.87\% & 17.39\% & 21.74\% & 0\%\\ \midrule
  Text chatbox & 3.32 & 22.73\% & 45.45\% & 18.18\% & 13.64\%\\ \midrule
  GUI & 3.24 & 23.81\% & 38.1\% & 38.1\% & 0\%\\ \midrule
  Virtual reality & 1.52 & 0\% & 6.25\% & 25\% & 68.75\%\\
\bottomrule
  \end{tabular}
\end{table}

\textbf{Opinion 6 (which agent ability is needed to develop):} In the future development of \mlas, ``more intelligent and autonomous decision-making capability'' is considered the most critical feature among our participants; almost half of the participants (47.83\%) rank it at first place.
The options ``Continuous improvement of user experience and interaction methods'' and ``Secure handling of personal data'' also received much attention, with 36.36\% and 33.33\% respectively, tying for the second place. 
Although "Integration with IoT devices" ranks last, 47.63\% of participants still believe it is important as an infrastructure for \mlas.

\textbf{Opinion 7 (what features are desired for an ideal IPA):}
Based on the responses from the participants, we summarize the following six key features of an ideal agent:
\begin{itemize}[left=1em]
\setlength{\itemsep}{0pt}
\setlength{\parsep}{0pt}
\setlength{\parskip}{0pt}
    \item \emph{Efficient Data Management and Search:} The agent acts as an external brain to remember the user's data by efficient data storage.
    It provides users with fast retrieval and precise search capabilities. 
    \item \emph{Work and Life Assistance:} The agent serves as a copilot in work when users ask for technical details. 
    It can also perform repetitive and heavy tasks and provide document and content generation for users. 
    \item \emph{Personalized Services and Recommendations:} According to user habits, the agent can discover the potential needs of users and then proactively provide services for users.
    It can serve as a personal and family health manager, medical server, shopping comparison assistance, travel assistance, etc.
    \item \emph{Autonomous Task Planning and Completion:} The agent can understand the user's intention, decompose the tasks proposed by the user and automatically perform them step by step (further in autonomous chain-of-thought functions), and help the user complete the steps that need manual with explicit instructions.
    \item \emph{Emotional Support and Social Interaction:} The agent can understand and help the user adjust their emotions by chatting.
    It can also understand users' relationships with different people, and help them write the response draft in users' voices.
    \item \emph{Digital Representative and Beyond:} The agent can represent the user to attend meetings, drive the car, go to work, and do any authorized tasks.
    It can truly understand the user and communicate and socialize with others in the present users themselves.
\end{itemize}

\textbf{Opinion 8 (what are the most urgent technical challenges):}
According to the responses from the participants, the most urgent challenges and technical issues are categorized as follows:
\begin{itemize}[left=0.5em]
\setlength{\itemsep}{0pt}
\setlength{\parsep}{0pt}
\setlength{\parskip}{0pt}
    \item \emph{Intelligence.} 1) \underline{Multimodal Support:} LLMs need to understand and process different data types (\eg text, images, and videos), thus it should possess advanced data alignment and interpretation capabilities.
    2) \underline{Context Understanding and Context-aware Actions:} In various application scenarios, LLMs must accurately understand user requirements and generate corresponding control instructions. This needs LLMs' context understanding ability and the ability to convert the context to effective actions.
    3) \underline{Enhancing Domain-specific Abilities of Lightweight LLM:} LLMs on resource-limited personal devices might underperform in complex tasks or understanding deep contextual meanings due to their size and complexity constraints. Therefore, how to boost the lightweight models' capabilities and handle complex tasks in specific domains is widely concerned.
    \item \emph{Performance.} 
    1) \underline{Effective LLM Compression or Compact Architecture:} Running LLMs on resource-limited mobile devices needs to balance the performance and quality of task completion.
    Efficient model compression techniques that concern the characteristics of LLMs to keep high quality of task completion are desirable.
    2) \underline{Practical Local-Remote Collaborative Architecture:} 
    Local-remote collaborative architecture of LLM is considered promising, which is desired to inherit both the fast/low-cost response ability of local model and the high-quality generation ability of the cloud model.
    However, how to achieve accurate and efficient collaboration is widely considered as an important challenge.
    \item \emph{Security \& Privacy.} 
    1) \underline{Data Security and Privacy Protection:} Ensuring the security of personal data and the protection of user privacy is critical when using personal data to train and execute LLMs. This proposes an urgent requirement to develop new data anonymization techniques and privacy protection protocols.
    2) \underline{Inference Accuracy and Harmlessness:} Ensure that the model outputs are precise and harmless for users, especially when used for decision-making or in sensitive scenarios.
    \item \emph{Personalization \& Storage.} Personalization requires efficient data storage solutions to manage and leverage user-related data, including their preferences, historical behaviors, and interactions.
    \item \emph{Traditional OS Support.} For mobile-based LLM agents, a critical requirement is LLM-friendly interfaces and support of traditional operating systems like Android. This may involve updates at the operating system level and the development of application programming interfaces (APIs) for better integration and utilization of LLM's functionalities.
\end{itemize}

Motivated by the valuable opinions of domain experts, the following sections will discuss the desired capabilities and potential challenges in more detail.

\section{Fundamental Capabilities}

\label{sec:fundamental-capabilities}

\begin{figure}[ht]
  \centering
  \includegraphics[width=13cm]{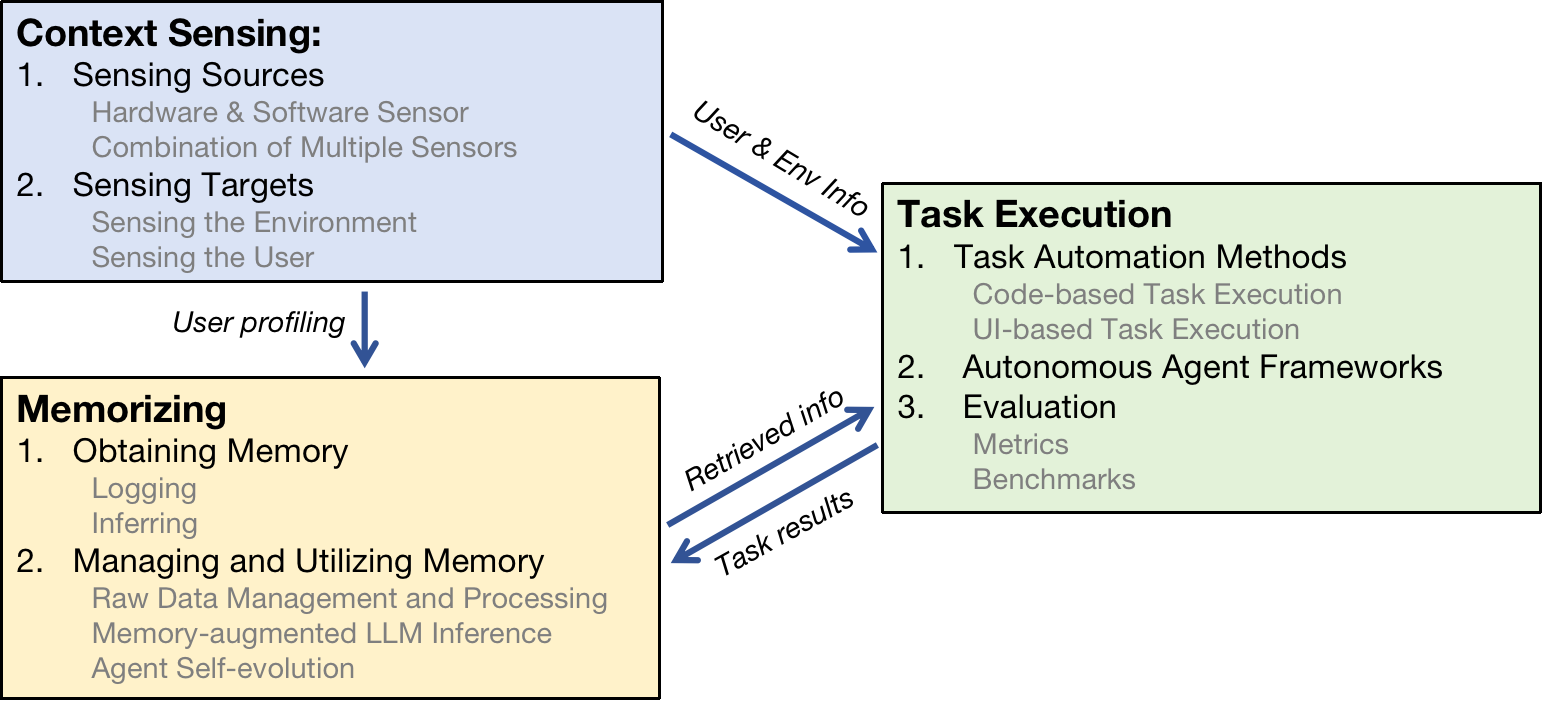}
  \caption{The fundamental capabilities of \mlas.}
  \label{figs:fund_cap}
\end{figure}

We first discuss the capabilities required by \mlas to support diverse features.
Excluding the general capabilities of normal LLM agents, we focus on three fundamental capabilities for personal assistants, including task execution, context sensing, and memorization. Task execution (\S\ref{sec: task_auto}) is to translate the users' commands or the proactively perceived tasks into actions on personal resources. The purpose of context sensing (\S\ref{sec: context_awa}) is to perceive the current state of the user and the environment, providing comprehensive information for task execution. Memorization (\S\ref{sec: mem}) is to record the user data, enabling the agent to recall past events, summarize knowledge and self-evolve. While context sensing and memorization are abilities associated with querying information from users, task execution refers to the ability of providing services to users. Figure \ref{figs:fund_cap} depicts the relation of these fundamental capabilities. The following sections discuss these capabilities in details.



\subsection{Task Execution}
\label{sec: task_auto}
Task execution is a fundamental capability of a \mla, enabling it to respond to user requests and carry out specified tasks. 
In our scenario, the agent is designed to interact with and control various personal devices such as smartphones, computers and IoT devices to automatically execute users' commands.

A fundamental requirement for task execution is the agent's ability to accurately interpret tasks as communicated by users. Typically, tasks may originate from users' verbal or written instructions, from which the intelligent agent discerns the user's intent. With the maturation of voice recognition technology, converting voice information into text has become highly convenient \cite{li2022speech, prabhavalkar2023endtoendspeech}. 

\mlas should make plans and take actions automatically after converting the users' commands into text. While planning poses a challenge for traditional DNNs, LLM-based agents exhibit greater proficiency in this regard. 
The planning and reasoning abilities of LLM agents have been discussed in the former surveys \cite{Gaoling_agent_survey, agent_survey, zhang2023igniting}. 
Our paper primarily focuses on the manipulation of personal data and interaction with personal devices. A significant consideration is that \mlas might need to interact with applications or systems that may lack comprehensive API support. 
Consequently, we also explore the user interface (UI) as an important tool for personal agents, enabling effective interaction in scenarios where API limitations exist.

\subsubsection{Task Automation Methods}
Based on the types of interaction mode, the methods of task execution can be categorized into code-based and UI-based approaches. In the code-based scenario, agents primarily complete tasks by automatically generating code to call APIs. Under UI-based scenarios, agents interact with personal devices by automatically simulating human interactions with the UI interface.

\textbf{Code-based Task Automation} often involves generating appropriate code to interact with APIs, databases, and DNN models. 
Traditional code-based personal assistants are often based on slot-filling-based task-oriented dialogue (TOD) frameworks. In the era of LLM, more researchers are attempting to directly use LLMs to directly generate code that calls APIs in order to accomplish more complex tasks. 
\begin{itemize}

    \item \textbf{Slot-filling method} is often used in task-oriented dialogue systems (TOD) or chatbots, which is conversational AI designed to assist users in completing specific tasks through dialogue \cite{POMDP-Based, rastogi-etal-2018-multi}. 
    In a task-oriented dialogue system, ``slots'' are predefined categories of information necessary to complete a task. 
    For example, in a travel booking application, slots might include destination, travel dates, number of passengers, etc. 
    During a conversation, the system prompts the user for this information, and calls corresponding APIs to complete the tasks.
    For mobile devices, many approaches focus on facilitating task automation by allowing users to demonstrate the desired tasks, which can be executed via a conversational interface \cite{kite, sugilite, li2017programming_iot, ulink}. These methods often assume that the user's tasks can be defined as a collection of slot-value pairs. 
    This assumption allows for precise management of the conversation with the controllable units, and to execute the task is to keep prompting users for the values of slots that have not been identified. 
    However, these methods do not consider complex cases where there are multiple values for a slot or relationships between slots \cite{TODinthewild}. Besides, they heavily rely on well-defined APIs and lack adaptability to unseen domains. Recent research papers utilize the understanding and reasoning ability of LLMs to complete more complex and multi-turn TOD tasks \cite{chung2023instructtods, hu2023enhancing, are_llms_tod, potential_user_feedback}, and improve the efficiency of Slot-filling methods.
    
    \item \textbf{Program synthesis method} is to utilize the code generation ability of LLMs to interact with APIs. 
    One way is to fine-tune LLMs to use specific APIs. 
    WebGPT \cite{nakano2022webgpt} fine-tunes a GPT-3 \cite{gpt3} to answer long-form questions by calling Microsoft Bing Web Search API \cite{bingapis}. 
    Some recent works \cite{schick2023toolformer, patil2023gorilla, yang2023gpt4tools, qin2023toolllm} fine-tune LLMs to retrieve and call APIs, enhancing their performance in various tasks like mathematical reasoning and program synthesis. 
    Octopus V2 \cite{chen2024octopus} introduces a 2B parameter on-device LLM to call Android APIs for task automation.
    Another way is to utilize the chain reasoning \cite{cot, tot, zhang2023igniting} and in-context learning ability \cite{gpt3} of LLMs. They show descriptions and demonstrations of the tools (e.g. APIs, other DNNs, etc.) in context and ask LLMs how to use them to complete tasks \cite{karpas2022mrkl, li2023camel, solve_computer, shen2023hugginggpt, lu2023chameleon}. 
    However, fine-tuning LLMs can be costly and restricted to the predefined set of tools, and in-context learning may fail when the number of APIs go large. Thus, authors of ToolkenGPT \cite{hao2023toolkengpt} attempt to solve this problem by representing each tool (API) as a token. 
    
\end{itemize}

Code-based methods can complete thousands of tasks from web searching to image generating. However, not all the needed APIs are available for agent developers in real-life apps out of security concerns or business interests. Besides, there are tasks that can be executed easily for human users but are difficult for calling system APIs \cite{TODinthewild}. Depending solely on publicly available APIs may not fully meet the highly diverse requirements for mobile task automation. 

\textbf{UI-based Task Automation.} Autonomous UI agents attempt to translate users' tasks into UI actions on smartphones or other personal devices, automating these tasks through direct UI interaction. Compared to code-based task execution, autonomous UI agents do not rely on publicly available APIs, potentially allowing for more versatile automation capabilities.
However, executing users' tasks by UI actions is not easy for traditional DNN models because of the implicit relations between tasks and UI elements. Recently, researchers utilize the comprehension and reasoning abilities of LLMs to improve the performance of autonomous UI agents.

The input of the UI agent is a task described in natural language, and a representation of the current UI, and the output is the UI action to be executed on the UI.
Depending on how they represent the UI, we can categorize the autonomous UI agents into text-based GUI representation and multimodal GUI representation.
\begin{itemize}

    \item \textbf{Text-based GUI representation} is to convert the UIs into pure text. 
    Seq2act \cite{seq2act} trains a transformer-based model \cite{transformer} to ground users' instruction to UI actions described in <operation, object, argument> tuples. 
    Researchers also investigate prompting with mobile UIs to complete tasks of UI instruction mapping \cite{talking_with_ui}. The authors convert mobile UI into HTML code, which is easy for LLMs to understand because an important part of their training data is scraped from Github. 
    DroidBot-GPT \cite{wen2023droidbotgpt} is an LLM-based system to complete users' tasks in a sequence of UI actions. Mind2Web \cite{deng2023mind2web} filters the raw HTML of webpages with a smaller LM and uses the LLM to select the target element and action.
    AutoDroid \cite{autodroid} uses app analysis tools to acquire app domain-specific knowledge and uses it to augment the LLMs for task automation. 
    In AXNav \cite{taeb2023axnav}, authors build a system using LLMs and pixel-based UI Understanding to execute manual accessibility tests. 
    MemoDroid \cite{memodroid} introduces an LLM-based mobile task automator that can break tasks into smaller sub-tasks and complete them by recalling former actions. 
    
    \item \textbf{Multimodal representation} is to use the image (and text) description of UI as the input of the \mlas. Early research work is focused on training multimodal transformers to ground user commands to UI elements \cite{metagui, actionbert, Li2023UINavAM}. In the era of LLMs, some approaches attempted to combine visual encoders with LLMs to handle GUI images \cite{responsible_task_automation, msra_ui2api, zhan2023autoui}. With the advent of Large Multimodal Models (LMMs), a growing number of projects employed visual language agents for UI action grounding and navigation \cite{from_pixel_to_actions, xie2024osworld}. One trend involves leveraging powerful LMMs such as GPT-4V to comprehend GUIs and select UI elements \cite{gpt4vWonderland, zhang2023appagent, gpt4v_web_agent, gao2023assistgui}. Another line of research is to customize open-sourced LMMs by fine-tuning on large-scale datasets for GUI-related tasks \cite{hong2023cogagent, cheng2024seeclick, you2024ferret}.
\end{itemize}

While UI-based task automation has the potential to achieve a more flexible personal agent framework compared to API-based automation, its research is still in the early stages. It remains challenging to accomplish more complex user commands. Besides, the privacy and security issues have not been fully addressed \cite{autodroid, responsible_task_automation}. It also remains controversial about the UI representation. While multimodal representation can handle elements that cannot be parsed through accessibility services, it is plagued by the heavy demands of screen recording and the limited reasoning abilities of current vision language models \cite{can_vlm_think}. 

\subsubsection{Autonomous Agent Frameworks}

An LLM-powered autonomous agent is composed of an LLM brain to make plans and self-reflection, a memory to store past information and knowledge, and a tool usage module to interact with tools (e.g. APIs, UIs, programming languages) \cite{LLMagentblog, agent_survey}.
There are a lot of popular projects that provide frameworks for users to create LLM-powered agents \cite{autogpt, langchain, babyagi, gptengineer, autoagents, OpenAgents, openinterpreter, Liu_LlamaIndex_2022, embedchain}. They attempt to enhance the ability of LLMs by interacting with other external tools and retrieving long/short-term memory. 
Auto-GPT \cite{autogpt} is one of the most famous frameworks, which can execute users' commands by generating prompts for GPT and using external tools.
LangChain \cite{langchain} is another popular framework that helps developers to create more sophisticated and context-aware applications using LLMs. 
Due to the ability to understand and produce natural language, LLM-powered agents can also engage with one another effortlessly, fostering an environment where collaboration and competition among multiple agents can thrive \cite{agents, hong2023metagpt, OpenAgents, wu2023autogen}. These autonomous agent frameworks make significant engineering contributions, providing a more user-friendly framework for the LLM-powered applications. 

For mobile devices, AutoDroid \cite{autodroid} provides an effective framework for developing mobile agents. Developers can easily create an automator for mobile tasks by either exploring apps using a test input generator or through manual demonstration. AutoDroid then automatically analyzes these records and utilizes them to improve Language Learning Models (LLMs) for more efficient task automation. 
\citet{automatic_macro_mining} develop a new method to effectively extract macros (basic units of user activity in apps such as ``login'', or ``call a contact'') from user-smartphone interaction traces. These macros can help agents to automatically complete tasks. 

\subsubsection{Evaluation} 

Evaluating the performance of task execution is a challenging issue. For API-based task execution, former surveys have provided a comprehensive summary on how to evaluate them \cite{Gaoling_agent_survey, zhang2023igniting}. Our paper mainly focuses on the evaluation of UI-based task automation. 

\textbf{Metrics: }
The metrics of UI-based task execution are completion rate \cite{seq2act, metagui, autodroid} and manually designed reward \cite{AndroidEnv, MobileEnv}. The completion rate is the probability that all actions predicted by the model are entirely consistent with the ground truth. However, since there may be different methods to complete a task, and the ground truth typically represents only one of these methods, the accuracy evaluated by this approach is not entirely correct \cite{autodroid}. Manually designing rewards based on the crucial steps can be more precise \cite{MobileEnv}, but they are less scalable because of the complex annotating process.

\begin{table}[ht]\small
\centering
\caption{UI task automation benchmarks. The structured UI form are view hierarchy (VH) and document object model (DOM) for Android and web respectively. For Windows, the metadata stems from the textual metadata within the operating system.}
\label{table:ui_benchmarks}
\begin{tabular}{c|c|c|c|c|c|c}
\toprule
Benchmark                        & Name                             & Platforms & \makecell{Human \\ annotations} & \makecell{UI \\ format} & \makecell{High-level \\ tasks} & \makecell{Exploration \\ memory}\\ \midrule
\multirow{10}{*}{Datasets}        & PhraseNode \cite{phrasenode}     & Web       & 51,663 & DOM, Screen & \ding{55} & \ding{55} \\ 
                                 & UIBert \cite{uibert}             & Web       & 16,660 & DOM, Screen & \ding{55} & \ding{55} \\ 
                                 & RicoSCA \cite{seq2act}           & Android   & N/A      & VH, Screen  & \ding{55} & \ding{55} \\ 
                                 & PixelHelp \cite{seq2act}         & Android   & 187    & VH, Screen  & \ding{51} & \ding{55} \\ 
                                 & MoTiF \cite{motif}               & Android   & 6,100  & VH, Screen  & \ding{51} & \ding{55} \\ 
                                 & META-GUI \cite{metagui}          & Android   & 4,684  & VH, Screen  & \ding{51} & \ding{55} \\ 
                                 & UGIF \cite{ugif}                 & Android   & 523    & VH, Screen  & \ding{51} & \ding{55} \\ 
                                 & Mind2Web \cite{deng2023mind2web} & Web       & 2,350  & DOM, Screen & \ding{51} & \ding{55} \\ 
                                 & AITW \cite{AITW_dataset}         & Android+Web   & 715,142& Screen      & \ding{51} & \ding{55} \\ 
                                 & DroidTask \cite{autodroid}       & Android   & 158    & VH, Screen  & \ding{51} & \ding{51} \\ 
                                 & OmniACT \cite{kapoor2024omniact} & Desktop+Web &9,802  & VH, Screen & \ding{51} & \ding{55} \\
                                 & AutoWebBench \cite{lai2024autowebglm} & Web &10,000  & DOM, Screen & \ding{51} & \ding{55} \\
                                  & VisualWebBench \cite{liu2024visualwebbench} & Web &1,500 & DOM, Screen & \ding{51} & \ding{55} \\
                                  & ScreenAgent \cite{niu2024screenagent} & Desktop &273 &  Screen & \ding{51} & \ding{55} \\
                                 \hline

\multirow{6}{*}{Platforms}    
                                 & MninWoB++ \cite{wob, Liu2018ReinforcementLO}    & Web       & 17,971 & DOM, Screen  & \ding{55} & \ding{51} \\ 
                                 & WebShop \cite{webshop}           & Web       & 12,087 & DOM, Screen  & \ding{51} & \ding{51} \\ 
                                 & WebArena \cite{zhou2023webarena} & Web       & 812    & DOM, Screen  & \ding{51} & \ding{51} \\ 
                                 & AndroidEnv \cite{AndroidEnv}     & Android   & N/A   & Screen   & \ding{51} & \ding{51} \\ 
                                 & MobileEnv \cite{MobileEnv}       & Android   & N/A     & VH, Screen   & \ding{51} & \ding{51} \\ 
                                 & AssistGUI \cite{gao2023assistgui}       & Windows   & 100     & Metadata, Screen   & \ding{51} & \ding{51} \\ 
                                 & OSWorld \cite{xie2024osworld} & Desktop &369  & VH, Screen & \ding{51} & \ding{51} \\
                                 & AgentStudio \cite{zheng2024agentstudio} & Desktop+Web &227  & DOM, Screen & \ding{51} & \ding{51} \\
                                 \bottomrule
                                 
\end{tabular}
\end{table}

\textbf{Benchmarks: }
Table \ref{table:ui_benchmarks} lists the benchmarks of UI-based task automation. One group of benchmarks is static datasets, which often include a set of human-annotated tasks, structured UI data (and screenshots), and actions to complete the tasks. Some of the tasks are synthetically generated \cite{seq2act, AndroidEnv, MobileEnv}. The early works mainly focus on low-level tasks with clear instructions \cite{phrasenode, uibert}, for example, \textit{click the `settings' button, and then click `Font size'}. Later works introduce high-level tasks that could be completed in multiple steps \cite{seq2act, motif, metagui, ugif, deng2023mind2web, AITW_dataset, kapoor2024omniact, lai2024autowebglm, liu2024visualwebbench, niu2024screenagent}, for example, \textit{delete all the events in my calendar}.
Another group of benchmarks are platforms that enable the agent to interact with. MiniWoB++ \cite{wob, Liu2018ReinforcementLO}, WebShop \cite{webshop}, and WebArena \cite{zhou2023webarena} provide web environments where agents can navigate and operate on the web by clicking, typing, closing page, and so on. AgentStudio \cite{zheng2024agentstudio} provides a comprehensive platform that supports interactions with versatile real-world computers. AndroidEnv \cite{AndroidEnv} and MobileEnv \cite{MobileEnv} provide a dynamic environment where agents can engage with any Android-based application and the core operating system. This framework allows for a wide scope of interaction and task-solving capabilities within the diverse Android platform. 

\begin{remark}

\textbf{Remark.} Existing approaches have demonstrated the remarkable ability of LLM agents in task reasoning and planning. However, there are several important problems to solve to realize practical \mlas.
\begin{enumerate}
  \item How to accurately and efficiently assess the performance of agents in real-world scenarios. 
  Because there are usually various ways to accomplish the same task, it is inaccurate to use a static dataset to measure the accuracy of task execution. Meanwhile, dynamically testing the tasks in a simulated environment may be inefficient and hard to reproduce.
  \item How to robustly determine if a task has been completed. LLMs often experience hallucinations during task execution, making it difficult to determine whether the current task has been completed. 
  \item Regarding UI agents, what is the best way to represent the software UI? The vision-based representation (e.g. screenshot) is generally available, while the text-based representation is usually more lightweight and friendly for LLM agents to operate. 
\end{enumerate}
\end{remark}


\subsection{Context Sensing}
\label{sec: context_awa}

Context Sensing refers to the process that the agent senses the status of the user or the environment, in order to provide more customized services.
In this work, we adopt a broad definition of context sensing, by considering generic information gathering process as a form of sensing. 
Hardware-based sensing aligns with the conventional notion of sensing, primarily involving data acquisition through various sensors, wearable devices, edge devices, and other data sources. 
On the other hand, software-based sensing emphasizes diverse means of data acquisition. For example, analyzing user typing habits and common phrases constitutes a form of software-base sensing. 

In \mlas, context sensing capability serves various purposes.
\textbf{1. Enabling Sensing Tasks}: Some tasks inherently require the agent to do sensing. For instance, when a user requires the agent to detect snoring during sleep, the agent must possess the ability to actively acquire, process, and analyze audio data.
\textbf{2. Supplementing Contextual Information}: The sensed information can facilitate the execution of ambiguous or complex tasks. For example, when the user wants to listen some music, it's good to know the current activity of the user to recommend appropriate music.
\textbf{3. Triggering Context-aware Services}: The sensing capability is also the basis to provide proactive services. For example, the agent may notice the users to keep focus upon detecting dangerous driving behaviors.
\textbf{4. Augmenting Agent Memory}: Some information perceived through sensing can become a part of the agent memory, which can be used by the agent for further customization and self-evolution. 

We introduce the techniques of context sensing from two perspectives, including sensing sources and sensing targets.

\subsubsection{Sensing Sources}

\textbf{Hardware Sensor.} Modern personal devices are equipped with a wide range of built-in hardware sensors, including accelerometers, gyroscopes, magnetic field sensors, light sensors, thermometers \cite{breda2023feverphone}, microphones \cite{chhaglani2022flowsense}, GPS modules, cameras \cite{hu2023microcam}, etc. Some other modules such as bluetooth and Wi-Fi \cite{hu2023muse} can also be used for sensing purposes. 
With the growing prevalence of wearable and IoT devices such as smart watches, bluetooth headphones \cite{gong2021robust}, and smart home devices \cite{arrotta2022dexar}, the sensing scope and sensing modalities are greatly expanded. 

Recently, there has been a proliferation of research exploring the deep integration of LLMs with raw sensor data. For instance, several studies directly embed raw IMU data into prompts for LLM, enabling Human Activity Recognition (HAR) \cite{ji2024hargpt} or trajectory prediction \cite{yang2024you}. \citet{zhang2024agent3d} provides LLM with a bird's-eye view of a 3D scene and allows it to iteratively select viewpoints to understand 3D point cloud scenes. Additionally, \citet{zheng2024bat} employs a trainable dual-channel audio frontend and fine-tuned LLM to enable LLM to comprehend spatial sound. Similar frontend and fine-tuning approaches are prevalent in various domains such as LiDAR \cite{yang2023lidar} and autonomous driving \cite{shao2023lmdrive, duan2024prompting}.

\textbf{Software Sensor.} Unlike hardware sensing that obtains data from real sensor devices, software sensing focuses on obtaining information from existing data, such as app usage \cite{wen2023towards}, call records \cite{bianchi2016identifying}, typing habits \cite{shin2023fedtherapist}, video game \cite{hu2024survey}, etc. The scope of software sensing is incredibly broad. For instance, in the field of natural language processing or audio, there exists a plethora of sensing research based on text or speech. Furthermore, recommendation systems such as e-commerce or short video platforms, the process typically involves first sensing certain user information and subsequently recommending specific products or content.
These sensors let agents better understand the users, enabling them to provide with more intelligent and personalized services.





\textbf{Combination of Multiple Sensors.}
Multi-sensor collaborative sensing stands out as an effective method for enhancing perceptual capabilities. Previous endeavors have demonstrated the assessment of user emotions, stress levels, and emotional states based on touchscreen and inertial sensors \cite{wampfler2022affective}, identification of time spent through screen capture and sensor data \cite{chen2023you}, breath detection through headphone microphones \cite{ahmed2023remote}, and nuanced motion detection through sensors and audio \cite{mollyn2022samosa}.

The significance of multi-sensor collaboration extends to the proliferation of intelligent wearables and smart homes. For instance, automatic recognition of when a user is working or resting using data collected from personal devices \cite{di2020multi} (smartwatches, laptops, and smartphones), or action detection through the combination of headphones and smartphone microphones \cite{gong2021robust}. Furthermore, technologies involving the fusion of household appliances, such as user action perception based on existing wired devices \cite{cui2023dancingant}, motion recognition in smart home environments \cite{arrotta2022dexar}, Wi-Fi-based motion detection \cite{he2023sencom}, multiperson detection \cite{hu2023muse}, and sleep monitoring \cite{zakaria2023sleepmore}.

There are three different approaches to enable LLM to understand and utilize sensor data.
\begin{itemize}
\item \textbf{Option 1: Sensor Data as Prompt.}
This method directly inputs sensor data into LLM as text prompts. Such an approach can be applied to various sensing sources such as IMU \cite{yang2024you} and bluetooth \cite{wang2024chattracer}. The mappings between the raw sensor data and the prompts can be created through rules, such as mapping tactile sensations on object surfaces to descriptors like ``soft'' or ``hard'' \cite{zhao2023chat}. This method is simple and effective as demonstrated many existing studies. However, it also has important limitations, such as the significant computational cost of processing large volumes of raw data and the limited ability of LLM to understand the complex sensor data in plain text.
\item \textbf{Option 2: Sensor Data Encoding + Fine-tuning.}
This approach enables LLM to understand sensor data with a data encoder. The encoder generates token embeddings from the raw sensor data with a learned neural network, and the embeddings are usually integrated into LLM through fine-tuning. This method yields significant results for complex sensor data, such as LiDAR \cite{yang2023lidar} and dual-channel audio \cite{zheng2024bat}. This approach allows LLM to efficiently understand sensor modalities, which is used to construct complex end-to-end systems like autonomous driving \cite{duan2024prompting, shao2023lmdrive}. Its drawback lies in the high training difficulty.
\item \textbf{Option 3: Redirecting Sensor Data to Domain-Specific Models.}
This approach doesn't process sensor data directly with LLMs, while it uses LLMs to invoke other specialized small models to deal with the raw sensor data. For example, \citet{darvish2024organa} leveraging techniques like object detection or pose estimation to assist chemical experiment robots in improving perception and understanding, additional information is added to the raw data stream and transformed into a form that LLMs can understand.

\end{itemize}

Multi-sensor and multi-device scenarios necessitate intricate considerations in data source selection, data fusion, and data analysis methods. Existing methodologies include LLM-driven strategies for generating multi-sensor policies in human behavior understanding \cite{gao2023automated}, emotion-agnostic multi-sensor data multitask learning frameworks \cite{samyoun2022m3sense}, cross-modal fusion of sensing data \cite{deldari2022cocoa}, wearable device motion recognition with a focus on multi-sensor fusion \cite{abedin2021attend}, and predictive anxiety in sensor data under conditions of data absence \cite{rashid2020predicting}. Furthermore, there are studies that analyze the importance of data features in fall detection \cite{kim2022fall}.


With the evolution of sensing technologies, multi-sensor and multi-device collaborative sensing has become a staple approach for perceiving complex scenarios. Effectively integrating diverse data sources to maximize accuracy and determining methods to eliminate less crucial data from a multitude of sources to conserve resources are vital research areas.

\subsubsection{Sensing Targets}

The objectives of context sensing can be categorized into environment sensing and user sensing. Environment sensing encompasses factors such as location, occasion, religious and cultural backgrounds, national and societal contexts, and more. Meanwhile, user sensing incorporates elements such as user activities, states, personal information, personality traits, emotions, goals, physical conditions, and other related aspects.

\textbf{Sensing the Environment.}
We further categorize environment sensing into two dimensions: \textit{scene sensing} and \textit{occasion sensing}. \textit{Scene sensing} predominantly involves more tangible environmental factors, such as locations and places. \textit{Occasion sensing} delves into deeper environmental information, including religious and cultural backgrounds, national differences, and social relationships.

\begin{itemize}

\item \textbf{Scene sensing} is often readily perceptible but hold significant importance, leading to variations both in behavior and emphasis. For behavior instance, detecting a user in a library prompts the agent to adjust the phone to silent mode, while in a bar increasing the volume and activating vibration may be necessary. Similarly to emphasis, when a user is in a meeting room, the agent should focus more on tasks related to meeting content recording and work organization, whereas in a gym, emphasis should shift towards fitness plans and heart rate analysis. Previous work in scene awareness has employed various techniques \cite{xu2023penetrative}, such as location-based approaches \cite{liu2013guoguo}, audio or video analysis \cite{chu2009environmental, chandrakala2019environmental}, and sensor capabilities analyzing aspects like airflow through smartphone microphones to assess ventilation \cite{chhaglani2022flowsense}, or scene recognition achieved by analyzing macro photographs taken with the smartphone camera when placed near a surface \cite{hu2023microcam}. \citet{zhang2024agent3d} let LLM understand 3D scenes through LLM-guided multiple viewpoint selection.

\item \textbf{Occasion perception} is more elusive in perception, and their impacts are relatively discreet. Earlier studies have identified differences in behavior and emotion recognition tasks across countries \cite{assi2023complex} and regions \cite{meegahapola2023generalization}. The national, ethnic, religious, and cultural backgrounds implied by the current user and setting are crucial. Perceiving others and objects in the current environment is equally vital. For example, previous work detected social scenarios based on sensor data, analyzing the behavior of socially anxious individuals in different social settings \cite{wang2023detecting}. Other research delved into analyzing drinking-related social scenes using multiple sensors, even predicting the size and gender composition of drinking groups \cite{meegahapola2021examining}. Additionally, studies explored the relations between sensor data, dietary habits, and social settings, revealing a strong association between binge eating and social environments, making it predictable \cite{meegahapola2021one}. \citet{liang2023exploring} use LLM forecasting pedestrian flow through the analysis of public events.

\end{itemize}


Environment sensing is crucial context information for a personal agent. Different environments lead to distinct behaviors and focal points, extending beyond mere locations to encompass social occasions, cultural backgrounds, and deeper conceptual elements, all environment individuals  and relationships, interactions, and anticipating the impacts on both the environment and the user. These considerations directly influence the level of intelligence exhibited by the personal agent.


\textbf{Sensing the User.}
User awareness is one of the primary features of \mlas. A deeper understanding of the user can better reflects the value and significance of the \mlas. We categorize user sensing into two temporal dimensions, including \textit{short-term} and \textit{long-term}. \textit{Short-term} sensing exhibits higher temporal variability and increased randomness. On the other hand, \textit{long-term} sensing necessitates extended maintenance and correction, making it relatively more stable and reliable.

\begin{itemize}
\item \textbf{Short-term user sensing} encompasses various aspects, including users' routine actions \cite{su2014activity}, or specialized activities such as tooth brushing effectiveness \cite{akther2021mteeth}, \citet{ji2024hargpt} found that even directly feeding IMU data to LLM can perform Human Activity Recognition (HAR) tasks. User states such as working or resting \cite{di2020multi, chen2023you}, user health conditions \cite{cao2022guard,breda2023feverphone,lin2020healthwalks}, as well as user emotions \cite{zhang2018moodexplorer, wampfler2022affective} and stress levels \cite{adler2021identifying}. Recently, numerous studies have attempted to explore the applications of LLMs in the field of health monitoring \cite{kim2024health, lan2024depression, lifelo2024adapting}. Short-term sensing typically involve rapidly changing and shallow-level state information. Efficiently capturing such information can significantly enhance the context awareness of \mlas.


\item \textbf{Long-term user sensing} mainly focus on the analysis of users' profile and personality. Various approaches have been proposed to understand users' work, study, and daily life. For instance, a study utilized sensor data from new smartphones to detect the prolonged psychological states of freshmen \cite{wang2022first}. Another study demonstrated the capability to predict learning performance and social activities based on perception data \cite{wang2015smartgpa}. \citet{gao2019predicting} delve into the techniques to predict personality based on the intensity of physical activities. 
There is also research examining the relationship between sensor data and user career advancement \cite{nepal2020detecting}, as well as a study that predicts user life satisfaction \cite{yuruten2014predictors}. Furthermore, specific states of users have been a focus, including studies on the perception of mental illnesses \cite{wang2020social,guo2024large}, such as one that predicts and analyzes schizophrenia \cite{wang2017predicting}, depression \cite{lan2024depression}, and another that detects habits like smoking \cite{chatterjee2020smokingopp}. \citet{lifelo2024adapting} utilized LLM to conduct psychological disorder analysis for a highly rare African language. Additionally, \citet{ouyang2024llmsense} attempt to extract higher-level perceptual information from simple data. Long-term sensing involve deep and abstract information, containing the profound logic behind user behavior. These pieces of information are often more subtle, making perception and maintenance challenging. However, they constitute an essential aspect for advanced personal agents.

\end{itemize}

In terms of user sensing, there are also several LLM-based initiatives, such as employing LLM for recommendation tasks \cite{chen2023palr, zhang2023bridging}, sentiment analysis with LLM \cite{sun2023sentiment}, and the development of a personal doctor equipped with inquiry and perception capabilities \cite{abbasian2023conversational}.


\begin{remark}

\textbf{Remark.} Existing methods often confine themselves to specific sensors, individual apps, or particular domains. In \mlas, a possible opportunity is to unify all sensing results concerning the environment and the user to originate from diverse sources. However, to achieve this goal involves several important research challenges.

\begin{enumerate}
  \item What is a unified format or ontology of the sensed information? The agents should be able to convert diverse sensing data into this format and conveniently use the data for various downstream tasks.
  \item Given the broad scope of sensing, how can the agents decide when and what to sense, in order to provide context-aware services with minimal overhead?
\end{enumerate}
\end{remark}

\subsection{Memorizing}
\label{sec: mem}


Memorizing denotes the capability to record, manage and utilize historical data in \mlas. This capability enables the agents to keep track of the user, learn from past experiences, extract useful knowledge, and apply this acquired knowledge to further enhance the service quality. 
The related work is mainly aimed to answer two questions, including how to obtain the memory and how to utilize the memory.

\subsubsection{Obtaining Memory}

The agent memory can be in various formats.
For example, the basic user profiles (\eg birthdate, addresses, personalities, preferences) are often stored in key-value pairs, allowing for easy key-based retrieval. 
Historical records are usually represented as sequences indexed by timestamps, which archive user service access, activities, system events and so on over the time. 
The user's documents, photos, videos, etc. are stored as files, which are often produced by other applications.
There are mainly two ways to obtain the memory: directly logging the raw data or indirectly inferring knowledge from raw data.

\textbf{Logging.} 
The most straightforward way to obtain memory is through logging, such as recording user input, system events, and sensed contexts. 
Logging data is often relatively simple. \emph{Life logging} is a commonly-discussed topic that focuses on tracking and recording user data created through the activities and behaviors of users, contributing to a comprehensive understanding of individuals' lifestyles and preferences \cite{gurrin2014lifelogging,dodge2007outlines}. Data recorded at specific moments using video cameras provide deeper overview of daily activities \cite{beddiar2020vision}. Moreover, recording data over long periods of time can provide valuable insights into behavior patterns, which will support the personalization of intelligent agents \cite{stachl2020predicting}. 

\textbf{Inferring.}
Another way of \mlas to obtain memory is to extract knowledge from the raw data. With the advancements in machine learning and data analytics, it has become possible to infer user behavior, patterns, and interactions to gain insights into their psychology, preferences, and other high-level information. For example, user personality can be extracted from texts \cite{majumder2017deep,vstajner2020survey}, emotions can be read from image and text data \cite{jaiswal2020facial,zad2021emotion}, preferences can be modeled from historical interaction information \cite{tang2019akupm}, and knowledge graphs can be extracted from smartphone push notifications \cite{li2018automated}. These extracted high-level information will also be stored as memories of the agent and utilized in services. 

\subsubsection{Managing and Utilizing Memory}

After obtaining the memory, the next question is how to manage and utilize the memory to provide better services in \mlas. Based on the purposes of utilizing memory, we divide the relevant techniques into following three parts, including raw data management, memory-augmented LLM inference, and agent self-evolution.

\textbf{Raw Data Management and Processing.}
A basic ability of \mlas is to access and process the raw memory data (\eg selecting, filtering, transforming to other formats, etc.), in order to facilitate other advanced functions. This line of work primarily focus on enabling more natural and human-comprehensible access, manipulation, and modification of data. Since the input-output and reasoning processes of LLMs are based on natural language, such interfaces are more easily integrated with other capabilities of large models. In this research area, numerous endeavors have explored the use of machine learning models or template-based methods to map user data requests to database SQL statements \cite{singh2016algorithm,lin2019grammar}. There are also framework-level works examining how to unify and simplify data interfaces. For instance, PrivacyStreams \cite{li_imwut17_privacystreams} unifies all personal data access and processing interfaces into a stream-based framework, which is more conducive for large language models to comprehend and manage.

\textbf{Memory-augmented LLM Inference.}
To enable the \mlas to provide customized services based on the user-related memory, it is usually desired to make use of the memory data in the LLM inference process. Recent research in LLM agents has explored leveraging memory to enhance decision-making and reasoning \cite{tot, park2023generative, li2023mot, wang2023augmenting, guozhicheng}, which provides inspiration for a solution where \mlas can offer personalized services to users through memories. The techniques can be different based on the types of the memory. 

\begin{itemize}
\item \textbf{Short-term memory} preserves and retains pertinent information in the form of symbolic variables, ensuring its accessibility and applicability during the current decision cycle. This includes perceptual inputs, active knowledge (generated by reasoning or retrieved from memory data), and other core information carried over from the previous decision cycle (\eg., agent's active goals). CoT \cite{cot}, Scratchpads \cite{nye2021show} encourage the LLM to generate intermediate reasoning, using the LLM's own context as a form of working memory. CoALA \cite{sumers2023cognitive} proposes that working memory should be a persistent data structure during long-term memory (LLM) calls. Each call generates its input from a subset of working memory (e.g., a prompt template and relevant variables), and the output is subsequently parsed into other variables (e.g., an action name and arguments) which are stored back in working memory and used to execute the corresponding action. 
In addition, short-term memory has the capability to interact with long-term memory and other data interfaces, serving as the central hub connecting different components of a language agent \cite{yao2022react,peng2023check}. 
\item \textbf{Long-term memory} stores experiences from earlier decision cycles. This can consist of history event flows \cite{park2023generative}, game trajectories from previous episodes \cite{tuyls2022multi,yao2020keep}, interaction information between the user and the agent or other representations of the agent's experiences. During the planning stage of a decision cycle, these episodes may be retrieved into working memory to support reasoning. An agent can also write new experiences from working to episodic memory as a form of learning. Secondly, long-term memory stores an agent's knowledge about the world and itself. Traditional approaches leverage retrieval for reasoning or decision-making initialize memory from an external database for knowledge support (\eg retrieval-augmented methods in NLP \cite{borgeaud2022improving,lewis2020retrieval}, ``reading to learn'' approaches in RL \cite{zhao2022process,hanjie2021grounding}). Agents may also write new knowledge obtained from LLM reasoning and user into long-term memory as a form of learning to incrementally build up world knowledge from experience.

\end{itemize}

\textbf{Agent Self-evolution.}
To better accommodate users, \mlas may also need to dynamically update themselves based on the memory data. We refer to this as ``self-evolution''.  The foundational functionality of intelligent agents is predominantly reliant on LLM. Therefore, the key to the self-evolution of intelligent agents lies in how to leverage LLM for the discovery and exploration of new skills, as well as in the continuous update of the LLM itself.

\begin{itemize}
  \item \textbf{Learning Skills.} Currently, numerous efforts are underway to enable LLM-based agents to engage in continuous skill learning and acquisition \cite{robotic,wang2023voyager}. These methods draw inspiration from the generality and  interpretability of programs \cite{ellis2023dreamcoder}, considering skills as executable code, and optimize skill acquisition by leveraging the in-context learning ability of LLM through the strategic use of prompts. They also manage a skill repository, integrating new skills as APIs, enabling intelligent agents to continually learn and reuse these skills in subsequent tasks. Prior work has demonstrated that modern LLMs can capture relevant information about meaningful skill chains \cite{huang2022language,singh2023progprompt}. Hence, intelligent agents have the capability to acquire novel skills by strategically linking skills within a foundational skill set \cite{zhang2023bootstrap}. In this process of skill chaining, the intelligent agent makes purposeful selections of subsequent meaningful skills, leveraging the a priori knowledge embedded in LLM and utilizing execution feedback to allow the language model to adjust its selections. This targeted approach enables the agent to efficiently assimilate complex skills.

  \item \textbf{Finetuning LLM.} To achieve the self-evolution of intelligent agents, continuous fine-tuning of the LLM is also required. There are several reasons: 1. Current LLMs were not specifically designed for agent-specific use cases, such as generating actions or self-evaluations, where limited learning support is provided by few-shot prompting. 2. Due to performance constraints on mobile devices, the capabilities of the LLM component of the intelligent agent are limited. This limitation makes it difficult for the model to acquire new skills through prior knowledge and in-context learning abilities. 3. During the operational phases of intelligent agents, the consistent emergence of materials such as the latest corpus \cite{jin2021lifelong}, new knowledge \cite{monaikul2021continual}, and tools \cite{qin2023tool} can frequently change the task schemas. This necessitates continual adaptation of LLMs. In such cases, fine-tuning the model becomes necessary to enhance its capacity for handling new tasks and generating appropriate actions. Research indicates that fine-tuned smaller LLMs could outperform prompted larger LLMs for specific reasoning \cite{zelikman2022star,huang2022large} and acting \cite{yao2022react} needs, while enjoying reduced inference time and expense. Parameter efficient fine-tuning (PEFT) \cite{houlsby2019parameter} presents a promising approach for efficiently fine-tuning LLMs. It only requires fine-tuning a small subset of external parameters \cite{peft}, making it friendly for edge devices, and it can effectively alleviate the issue of catastrophic forgetting \cite{wang2022adamix}. There have also been some preliminary attempts to conduct the study of LLM fine-tuning for agents \cite{chen2023fireact} with trajectories from multiple tasks and prompting methods, inspiring future endeavors aimed at developing more capable and useful \mlas.
\end{itemize}

\begin{remark}

\textbf{Remark.} The ability to generate and leverage the memory about the user is the basis of personalization in \mlas. We highlight following three open problems surrounding the memory mechanism of \mlas.

\begin{enumerate}
  \item The agent memory can potentially be huge, heterogeneous and dynamic. What is the most effective and efficient way for the agents to organize and retrieve the memory?
  \item Human has the ability to forget. Since inappropriate data in the memory can be harmful for the agents' service quality and efficiency, how can the agents determine what information to memorize?
  \item What is the best way for the agents to self-evolve with the memory? Specifically, what data to use, when to evolve, and how (fine-tuning or else)? How can the personalized models accept updates of the base foundation model?
\end{enumerate}
\end{remark}

\section{Efficiency}

\label{sec:efficiency}

\begin{figure}[htbp]
    \centering
    \includegraphics[width=.8\linewidth]{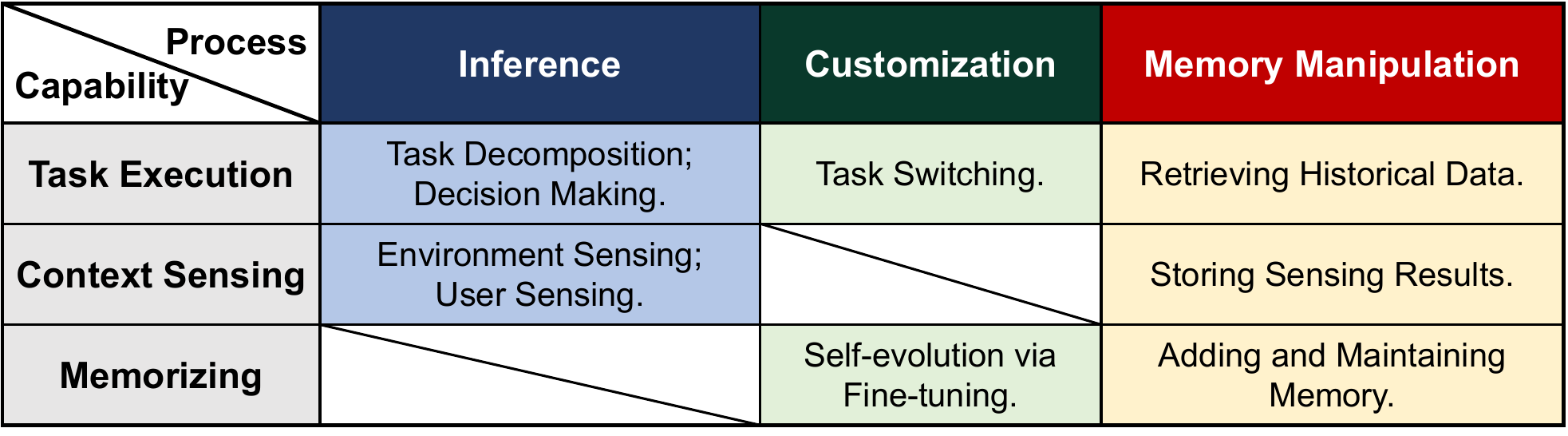}
    \caption{The mapping relations between the low-level processes and high-level capabilities of \mlas.}
    \label{fig:capability-decomposition}
\end{figure}

Due to the limited hardware resource and power supply on many personal devices, it is important to improve the efficiency of \mlas in the deployment stage.
We've discussed in Section~\ref{sec:fundamental-capabilities} the fundamental capabilities of \mlas, including task execution, context sensing, and memorizing.
These capabilities, as shown in Figure~\ref{fig:capability-decomposition}, are backed by more elementary processes, mainly including the inference, customization and memory retrieval of the LLM agent.
Each of these processes desires careful optimization of efficiency, as described below.

\textbf{Inference} of LLMs is the basis of an agent's various capabilities.
For example, the agent may first decompose a complex task into several steps with the help of the LLM, then solve each step through either LLM inference or invoking personal tools (\eg schedule a meeting).
Sensing the context or generating the memory may also rely on the reasoning abilities of LLMs.
While the cost of using the tools or sensors is usually hard to estimate due to the diversity, LLM inference is a common procedure that demands a lot of both computation and memory resources.
Therefore, the LLM inference becomes the performance bottleneck for the \mlas, requiring careful optimizations on its efficiency.


\textbf{Customization} is another important process of \mlas for accommodating different user requirements.
Customization is needed when the agents are installed to different users or used in different scenarios. The self-evolution of \mlas is also a process of customization.
To offer customized services, an agent can either feed the LLM with different context tokens or tune the LLM with domain-specific data.
Due to the frequent needs of customization, the processes may impose considerable pressure on the system's computational and storage resources.

\textbf{Memory manipulation} is another costly process. To provide better services, the agents may require access to longer contexts or external memories, such as environment perceptions, user profiles, interaction histories, data files, etc. Consequently, this gives rise to two considerations. The first pertains to necessitating LLMs to handle longer inputs. The second issue centers around the management and acquisition of information from an external memory bank.

\tikzstyle{my-box}=[
rectangle,
draw=hidden-draw,
rounded corners,
text opacity=1,
minimum height=1.5em,
minimum width=5em,
inner sep=2pt,
align=center,
fill opacity=.5,
line width=0.8pt,
]
\tikzstyle{leaf}=[my-box, minimum height=1.5em,
fill=hidden-pink!80, text=black, align=left,font=\normalsize,
inner xsep=2pt,
inner ysep=4pt,
line width=0.8pt,
]

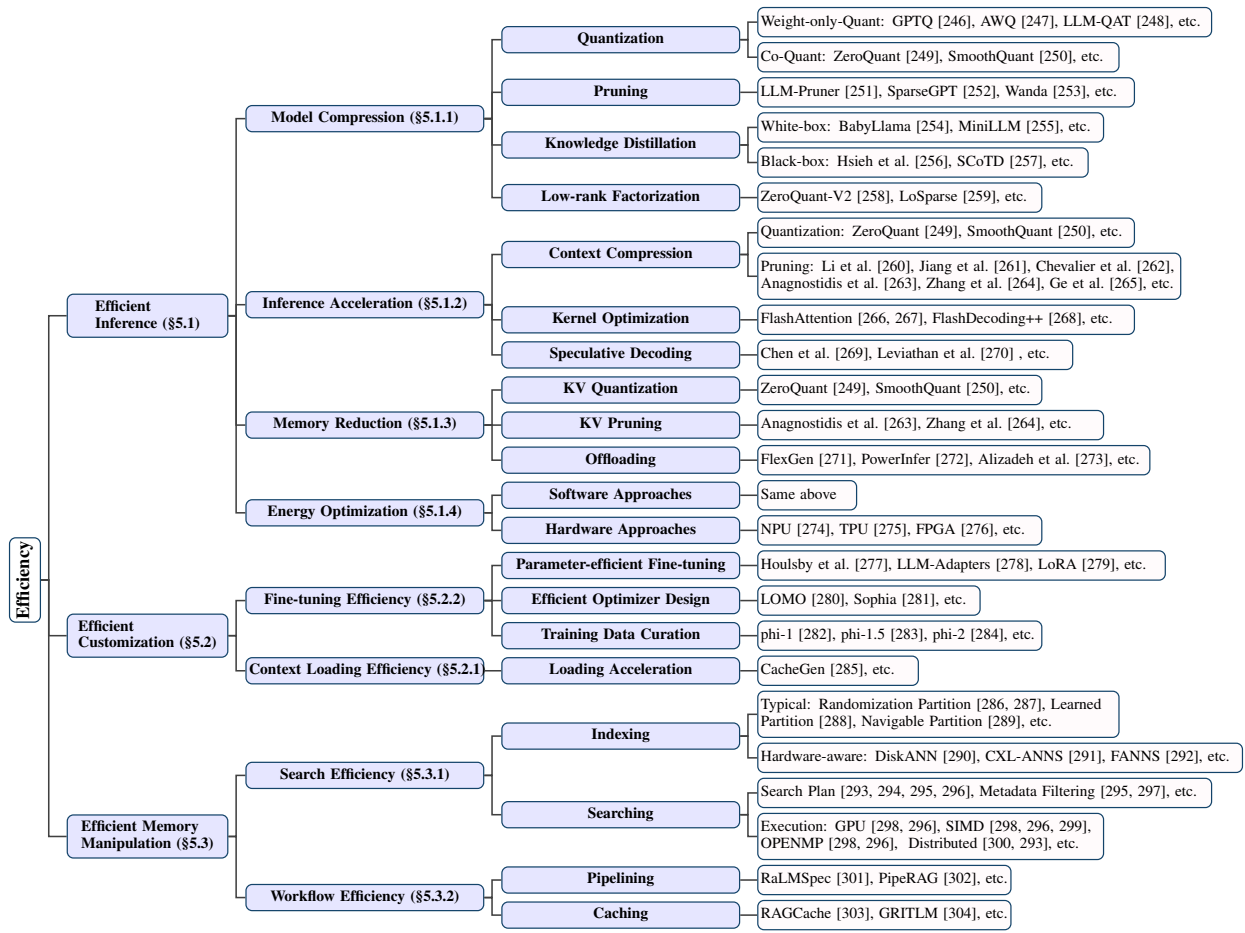
\begin{figure*}[t!]
    \centering
    \resizebox{\textwidth}{!}{
        \begin{forest}
            forked edges,
            for tree={
            grow=east,
            reversed=true,
            anchor=base west,
            parent anchor=east,
            child anchor=west,
            base=center,
            font=\large,
            rectangle,
            draw=hidden-draw,
            rounded corners,
            align=left,
            text centered,
            minimum width=4em,
            edge+={darkgray, line width=1pt},
            s sep=3pt,
            inner xsep=2pt,
            inner ysep=3pt,
            line width=0.8pt,
            ver/.style={rotate=90, child anchor=north, parent anchor=south, anchor=center},
            },
            where level=1{text width=10em,font=\normalsize,}{},
            where level=2{text width=15em,font=\normalsize,}{},
            where level=3{text width=15em,font=\normalsize,}{},
            [\textbf{Efficiency}, ver
            [\textbf{Efficient} \\ \textbf{Inference (\S\ref{sec:inference-efficiency})}, fill=blue!10
            [\textbf{Model Compression (\S\ref{sec:generic-efficiency})}, fill=blue!10
            [\textbf{Quantization}, fill=blue!10
            [
            Weight-only-Quant: GPTQ~\cite{frantar2022gptq}{, }AWQ~\cite{lin2023awq}{, }LLM-QAT~\cite{liu2023llm-qat}{, }etc. , leaf, text width=29em
            ]
            [
            Co-Quant: ZeroQuant~\cite{yao2022zeroquant}{, }SmoothQuant~\cite{xiao2023smoothquant}{, }etc. , leaf, text width=23em
            ]
            ]
            [\textbf{Pruning}, fill=blue!10
            [
            LLM-Pruner~\cite{ma2023llm-pruner}{, }SparseGPT~\cite{frantar2023sparsegpt}{, }Wanda~\cite{sun2023simplepruning}{, }etc. , leaf, text width=24em
            ]
            ]
            [\textbf{Knowledge Distillation}, fill=blue!10
            [
            White-box: BabyLlama~\cite{timiryasov2023baby-llama}{, }MiniLLM~\cite{gu2023minillm}{, }etc. , leaf, text width=22em
            ]
            [
            Black-box: \citet{hsieh2023distilling-step}{, }SCoTD~\cite{li2023symbolic-chain}{, }etc. , leaf, text width=21em
            ]
            ]
            [\textbf{Low-rank Factorization}, fill=blue!10
            [
            ZeroQuant-V2~\cite{yao2023zeroquantv2}{, }LoSparse~\cite{li2023losparse}{, }etc. ,leaf, text width=18em
            ]
            ]
            ]
            [\textbf{Inference Acceleration (\S\ref{sec:computational-efficiency})}, fill=blue!10
            [\textbf{Context Compression}, fill=blue!10
            [
            Quantization: ZeroQuant~\cite{yao2022zeroquant}{, }SmoothQuant~\cite{xiao2023smoothquant}{, }etc. , leaf, text width=24em
            ]
            [
            Pruning:
            \citet{li2023compressing-context}{, }\citet{jiang2023llmlingua}{, }\citet{AdaptingLM}{, }\\
            \citet{anagnostidis2023dynamic-context-pruning}{, }\citet{zhang2023h2o}{, }\citet{ge2024model}{, }etc. , leaf, text width=27em
            ]
            ]
            [\textbf{Kernel Optimization}, fill=blue!10
            [
            FlashAttention~\cite{dao2022flashattention, dao2023flashattention}{, }FlashDecoding++~\cite{hong2023flashdecoding++}{, }etc. , leaf, text width=24em
            ]
            ]
            [\textbf{Speculative Decoding}, fill=blue!10
            [
            \citet{speculative-chen2023accelerating}{, }\citet{speculative-leviathan2023fast}
            {, }etc. , leaf, text width=20em
            ]
            ]
            ]
            [\textbf{Memory Reduction (\S\ref{sec:memory-efficiency})}, fill=blue!10
            [\textbf{KV Quantization}, fill=blue!10
            [
            ZeroQuant~\cite{yao2022zeroquant}{, }SmoothQuant~\cite{xiao2023smoothquant}{, }etc. , leaf, text width=18em
            ]
            ]
            [\textbf{KV Pruning}, fill=blue!10
            [
            \citet{anagnostidis2023dynamic-context-pruning}{, }\citet{zhang2023h2o}{, }etc. , leaf, text width=22em
            ]
            ]
            [\textbf{Offloading}, fill=blue!10
            [
            FlexGen~\cite{sheng2023flexgen}{, }PowerInfer~\cite{song2023powerinfer}{, }\citet{alizadeh2023llm-in-a-flash}{, }etc. , leaf, text width=25em
            ]
            ]
            ]
            [\textbf{Energy Optimization (\S\ref{sec:energy-efficiency})}, fill=blue!10
            [\textbf{Software Approaches}, fill=blue!10
            [
            Same~above, leaf, text width=6em
            ]
            ]
            [\textbf{Hardware Approaches}, fill=blue!10
            [
            NPU~\cite{qualcomm_npu}{, }TPU~\cite{reidy2023efficient-tpu}{, }FPGA~\cite{hong2022dfx-fpga}{, }etc. , leaf, text width=18em
            ]
            ]
            ]
            ]
            [\textbf{Efficient} \\ \textbf{Customization (\S\ref{sec:customization-efficiency})}, fill=blue!10
            [\textbf{Fine-tuning Efficiency (\S\ref{sec:fine-tuning-efficiency})}, fill=blue!10
            [\textbf{Parameter-efficient Fine-tuning}, fill=blue!10
            [
            \citet{houlsby2019parameterefficient}{, }LLM-Adapters~\cite{hu2023llmadapters}{, }LoRA~\cite{hu2022lora}{, }etc. , leaf, text width=26em
            ]
            ]
            [\textbf{Efficient Optimizer Design}, fill=blue!10
            [
            LOMO~\cite{lv2023parameter}{, }Sophia~\cite{liu2023sophia}{, }etc. , leaf, text width=14em
            ]
            ]
            [\textbf{Training Data Curation}, fill=blue!10
            [
            phi-1~\cite{gunasekar2023textbooks}{, }phi-1.5~\cite{li2023textbooks}{, }phi-2~\cite{li2023phi2}{, }etc. , leaf, text width=18em
            ]
            ]
            ]
            [\textbf{Context Loading Efficiency (\S\ref{sec:context-loading-efficiency})}, fill=blue!10
            [\textbf{Loading Acceleration}, fill=blue!10
            [
            CacheGen~\cite{liu2023cachegen}{, }etc. , leaf, text width=10em
            ]
            ]
            ]
            ]
            [\textbf{Efficient Memory} \\ \textbf{Manipulation (\S\ref{sec:memory-retrieval-efficiency})}, fill=blue!10
            [\textbf{Search Efficiency (\S\ref{sec: Efficient Search})}, fill=blue!10
            [\textbf{Indexing}, fill=blue!10
            [
            Typical: Randomization Partition \cite{E2LSH,RPTree1}{, }Learned \\Partition \cite{spann}{, }Navigable Partition \cite{HNSW}{, }etc. , leaf, text width=23em
            ]
            [
            Hardware-aware: DiskANN \cite{Subramanya2019DiskANNFA}{, }CXL-ANNS \cite{CXL-ANNS}{, }FANNS~\cite{FANNS}{, }etc. , leaf, text width=31em
            ]
            ]
            [\textbf{Searching}, fill=blue!10
            [
            Search Plan \cite{qdrant, vespa, AnalyticDB-V, Milvus}{, }Metadata Filtering \cite{AnalyticDB-V, HQANN}{, }etc. , leaf, text width=29em
            ]
            [
            Execution: GPU~\cite{faiss, Milvus}{, }SIMD \cite{faiss, Milvus, quickeradc}{, }\\OPENMP \cite{faiss, Milvus}{, } Distributed~\cite{vald,qdrant}{, }etc. , leaf, text width=22em
            ]
            ]
            ]
            [\textbf{Workflow Efficiency (\S\ref{sec: Efficient Workflow})}, fill=blue!10
            [\textbf{Pipelining}, fill=blue!10
            [
            RaLMSpec \cite{Zhang2024AcceleratingRL}{, }PipeRAG~\cite{jiang2024piperag}{, }etc. , leaf, text width=16em
            ]
            ]
            [\textbf{Caching}, fill=blue!10
            [
            RAGCache \cite{jin2024ragcache}{, }GRITLM \cite{muennighoff2024generative}{, }etc. , leaf, text width=16em
            ]
            ]
            ]
            ]
            ]
        \end{forest}
    }
    \caption{Overview of techniques to improve the efficiency of LLM agents. The leaf nodes are part of representative works we have cited.}
    \label{fig:efficiency-overview}
\end{figure*}

We'll dive into the efficiency of each component in the following subsections, as is shown in Figure~\ref{fig:efficiency-overview}.

\subsection{Efficient Inference}
\label{sec:inference-efficiency}

Since the runtime cost of \mlas is dominated by LLM inference, it is important to improve the inference efficiency to enhance the overall efficiency of the agent.
Although the total inference cost can be significantly influenced by the design of agents, including how the agents send requests to LLMs, what prompts to use, etc., we will be focused on model and system-level approaches only.
The reason is that the designs of agents may vary based on the actual applications and don't directly contribute to the efficiency of LLM inference itself.

Many model and system-level approaches have been proposed to improve the efficiency of LLM inference. While some of them are generic for the overall performance and efficiency (\eg model compression), there are also techniques targeting the efficiency of specific perspectives, such as model size, inference latency, memory consumption, energy consumption, etc. We will discuss these aspects separately in the following parts of this subsection.

\subsubsection{Model Compression}

\label{sec:generic-efficiency}

Model compression techniques, which directly reduce the model size and computations, are generic optimizations to enhance the inference efficiency of LLMs, including computation, memory, energy and etc.
The model compression techniques are further categorized into various approaches, including quantization, pruning (sparsity), distillation and low-rank factorization.

\textbf{Quantization} is one of the most important compression approaches for LLMs.
It reduces the model size by using fewer bits to represent the model parameters, and also reduces computations with system-level support for quantized kernels.
Quantization methods can be further divided into post-training quantization (PTQ) and quantization-aware training (QAT), based on whether additional training is required after quantization.
Unlike QAT (\eg LLM-QAT~\cite{liu2023llm-qat}) which requires non-negligible additional training effort, PTQ is more available and flexible for on-device deployment under different hardware constraints.

Recent works have revealed that the difficulty of LLM quantization mainly lies in activations, where the outliers are hard to quantize~\cite{bondarenko2021understanding,wei2022outlier}.
Existing works have proposed various approaches to tackle this challenge.
A typical line of work adopts the weight only quantization (WOQ) paradigm, which conduct integer quantization (\eg INT4 and INT8) on weights only, while preserving activations in float formats (\eg FP16 and FP32). WOQ achieves a trade-off between the compression ratio and model perplexity.
A straightforward way of WOQ is the group-wise uniform quantization implemented in current mobile deployment frameworks (\eg llama.cpp~\cite{llama.cpp} and MLC-LLM~\cite{mlc-llm}).
Recent works also proposed different quantization algorithms to enhance model capability, such as GPTQ~\cite{frantar2022gptq} and AWQ~\cite{lin2023awq}.

Despite the WOQ techniques, another line of work quantizes both weights and activations.
For example, ZeroQuant~\cite{yao2022zeroquant} performs INT8 quantization for both weights and activations, using group-wise quantization for model weights and token-wise quantization for activations.
However, the activations, including key-value (KV) pairs, are usually more difficult to quantize compared to model weights because of outliers.
There have been extensive works to tackle this challenge.
SmoothQuant~\cite{xiao2023smoothquant} migrates the quantization difficulty of activations to weights through additional scaling operations that ``smooth'' the outliers in activations, and thereby achieve negligible accuracy degradation in W8A8 quantization.
Subsequent works further attempt to lower the usable quantization bitwidth down to 4-bit through various techniques including channel re-ordering (RPTQ~\cite{yuan2023rptq}), channel-wise shifting and scaling (Outlier Suppression+~\cite{wei2023outlier}), and adaptive channel reassembling (QLLM~\cite{liu2023qllm}).
Notably, RPTQ addresses the KV storage issue by developing a new quantization scheme that focuses solely on KV cache when quantizing activations, which is the major memory consumer in long-context inference.

While integer quantization methods such as INT4 and INT8 remain mainstream solutions in current deployment practice, there has been a new trend of low-bit floating point quantization, such as FP4 and FP8. One reason is that floating point quantization can achieve comparable or even higher accuracy than integer quantization \cite{zhang2023integer-or-floating-point, wu2023zeroquant-fp, liu2023llm-fp4}. Besides, floating point quantization is possible to achieve higher computational performance on both cloud GPUs like NVIDIA H100 with dedicated computing support, and mobile GPUs \cite{li2024transformer-lite}.

\textbf{Pruning} reduces the model size and computations by removing less important connections in the network.
Pruning is categorized into structured pruning and unstructured pruning.
Structure pruning usually removes weights in regular patterns, such as a rectangle block in the matrix or an entire channel, while unstructured pruning doesn't impose such constraints.
Consequently, structured pruning (\eg LLM-Pruner~\cite{ma2023llm-pruner}) is more hardware-friendly but more difficult to maintain model accuracy.
While traditional pruning approaches require costly retaining process to preserve model capability, recent works like SparseGPT~\cite{frantar2023sparsegpt} and Wanda~\cite{sun2023simplepruning} have explored to perform unstructured or semi-structured pruning in one-shot. 

\textbf{Knowledge Distillation (KD)}
involves using a well-performing teacher model (usually with a large number of parameters and high precision) to guide the training of a lightweight student model (usually with fewer parameters and lower precision).
Through distillation, the student model is well-aligned to the teacher model with relative smaller training dataset, and has the chance to perform even better on downstream tasks~\cite{hsieh2023distilling-step}.
Based on whether the teacher model's parameters are required in the training process, distillation methods can be further categorized into white-box (\eg BabyLlama~\cite{timiryasov2023baby-llama} and MiniLLM~\cite{gu2023minillm}) and black-box ones (\eg Distilling Step-by-Step~\cite{hsieh2023distilling-step} and SCoTD~\cite{li2023symbolic-chain}).
Since the student model are often lightweight quantized or pruned model, KD is also adopted in QAT and pruning techniques to enhance the training performance.
For example, LLM-QAT~\cite{liu2023llm-qat} proposes a data-free distillation method to preserve the original output distribution in the quantized model.

\textbf{Low-rank Factorization}
refers to approximating the original weight matrix by the product of two low-rank matrices, thereby reducing the model's parameter size and computational load.
Specifically, a weight matrix $W$ of shape $m\times n$ is factorized into the product of $U^{m\times r}$ and $V^{n\times r}$, such that $W\approx UV^T$ and $r\ll m,n$.
Low-rank Factorization can be combined with quantization (\eg ZeroQuant-V2~\cite{yao2023zeroquantv2}) and pruning (\eg LoSparse~\cite{li2023losparse}) methods to enhance the compression ratio.
Besides, low-rank adapters effectively reduce the customization overhead of LLMs, which we leave to \ref{sec:customization-efficiency}.

\subsubsection{Inference Acceleration}

\label{sec:computational-efficiency}

Except for making the models more compact as discussed in Section~\ref{sec:memory-efficiency}, there are various other techniques to accelerate the LLM inference process.

A major characteristic that sets the LLM apart from the traditional non-Transformer models is the attention mechanism \cite{transformer}.
Since the computational cost of attention increases near quadratically with the context length, it is particularly important to enhance the computational efficiency of long-context inference.
Existing works have explored to reduce context length and optimize attention kernels to better support long-context inference.
We'll dive into these techniques separately.

\textbf{KV Cache} is a widely adopted technique in both mobile (\eg llama.cpp~\cite{llama.cpp} and mlc-llm~\cite{mlc-llm}) and cloud LLM serving frameworks (\eg DeepSpeed~\cite{aminabadi2022deepspeed-inference} and vLLM~\cite{kwon2023vLLM}), to avoid redundant computation in LLM inference.
Specifically, KV Cache involves storing (\ie ``caching'') and incrementally updating the Key-Value (KV) pairs, which are intermediate results in the attention calculation, in each token's generation.
Therefore, the repeated part in the KV computation is avoided to reduce the computational cost.
However, in long-context inference, the computational cost of attention is still a system bottleneck despite the skipped KV calculations, making it crucial to compress the context length in such scenarios.

\textbf{Context Compression} methods enhance the inference efficiency by reducing the length of the context, especially the KV cache.
Co-quantization of weights and activations, including KV cache, is an intuitive approach to compress the KV cache, which has been discussed in Section~\ref{sec:generic-efficiency}.
Besides quantization, context pruning removes less important tokens in the context to reduce the computational cost.
The effectiveness of this method is based on the observation that tokens have different impacts on the final output, and removing less important tokens won't cause significant degradation of the model's capability~\cite{anagnostidis2023dynamic-context-pruning, liu2023scissorhands, zhang2023h2o, ge2024model}.
A typical line of work is to compress the context at the prefill stage based on different importance of tokens~\cite{li2023compressing-context, jiang2023llmlingua, AdaptingLM}.
However, these methods are one-shot and cannot prune the KV cache when the context length continuously grows during token generation.
To address this, Dynamic Context Pruning~\cite{anagnostidis2023dynamic-context-pruning} uses a learnable mechanism to continuously determine and drop uninformative tokens.
While the learnable mechanism introduces a fine-tuning overhead, \citet{zhang2023h2o}, proposes a token eviction strategy that can be applied without fine-tuning.

Inspired by the same observation that tokens are not equally important, other works also explored to reduce computations of less important tokens instead of directly removing them.
COLT5~\cite{CoLT5} employs a conditional computation mechanism, which devotes more resources to important tokens in both FFN and attention.
SkipDecode~\cite{del2023skipdecode} designs a token-level early exit method that works seamlessly with batched inference and KV cache, to skip some operators in the computational graph when a token is less important.

\textbf{Kernel Optimization} is another approach towards LLM inference acceleration.
Optimization for small-batch or single-batch inference is especially important for edge scenarios including the locally-deployed \mlas.
Existing works have revealed that the attention calculation becomes a bottleneck when the sequence length is long, since the complexity of attention scales quadratically with the sequence length, while that of the FFN scales linearly.
Therefore, efficient attention kernels including FlashAttention~\cite{dao2022flashattention,dao2023flashattention} and FlashDecoding++~\cite{hong2023flashdecoding++} have been proposed to improve the speed of long-text inference.
Some works also reduce the computational complexity of attention from the algorithm aspect.
For example, Linformer~\cite{Linformer} achieves linear complexity for self-attention in the prefill phase.
Besides, reducing dequantization overhead also provides significant performance improvement as demonstrated by LUT-GEMM~\cite{park2022lut-gemm}.



\textbf{Speculative Decoding}~\cite{speculative-leviathan2023fast,speculative-chen2023accelerating} is an effective approach in small-batch inference to improve the latency.
The batch size of LLM inference at the edge is smaller than on the cloud, and is usually 1 (\ie single query), which makes the inference workload extremely memory-bound.
Speculative decoding mitigates this challenge by ``guessing'' several subsequent tokens through a lightweight ``draft model'', and then validating the draft tokens in batches using the large ``oracle model''.
\citet{miao2023specinfer} and \citet{speculative-spector2023accelerating} further enhance speculative decoding with a tree-based verification instead of sequential ones to reuse intermediate results shared across these sequences.
While these methods ensure zero bias in the generated results, BiLD~\cite{kim2023speculative} proposes to only fallback or rollback to the oracle model occasionally when the draft model is not capable to generate high quality contents.

\subsubsection{Memory Reduction}

\label{sec:memory-efficiency}

LLM inference is not only computationally-intensive, but also memory-consuming, which causes challenges in the deployment of \mlas.
Therefore, it is necessary to perform optimizations on the memory efficiency of LLM inference.
KV cache and model weights are two major causes of this memory overhead.
In a short-context scenario where the KV storage requires much less memory than the model weights, the model compression techniques in Section~\ref{sec:generic-efficiency} are very effective to reduce the memory requirement to store the weights.
However, in the long-context scenario, the KV cache, whose size grows linearly with the context length, will dominate the total memory consumption.

An effective approach to address this issue is to \textbf{compress the KV cache} using quantization and pruning techniques mentioned in Section~\ref{sec:generic-efficiency} and Section~\ref{sec:computational-efficiency}.
While the quantization methods are generic to reduce the memory footprint of KV cache, not all the pruning-based methods directly contribute to the memory efficiency.
Only those methods that prune the corresponding rows/columns in the KV cache when continuously removing input tokens in the context can prevent the KV cache size from exceeding the memory limit.
For example, \citet{anagnostidis2023dynamic-context-pruning} and \citet{zhang2023h2o} proposed to identify and evict uninformative tokens during generation.
However, the one-shot approaches that only prunes the context at prefill stage are less effective regarding the generative scenarios.

Although the compression-based methods are demonstrated to be able to effectively reduce the memory requirement of LLM inference, the accuracy degradation caused by compression are not negligible in some cases. To address this, FlexGen~\cite{sheng2023flexgen} designs an offloading strategy to fully utilize GPU, CPU and disk, together with a zig-zag scheduling scheme to support high-throughput inference under constrained GPU memory. This approach is orthogonal to compression-based methods, and thus can be jointly used to further reduce GPU memory footprints.
Another line of work, including PowerInfer~\cite{song2023powerinfer} and \citet{alizadeh2023llm-in-a-flash}, reduces swapping overhead in low-batch inference by predicting contextual sparsity as inspired in \cite{liu2023deja-vu}.

\subsubsection{Energy Optimization}

\label{sec:energy-efficiency}

The energy consumption is a critical factor that affects the real-world deployment of LLM agents given LLM's costly computations and memory accesses. 
An energy-consuming agent not only increases the runtime cost and carbon footprint, but also hurts the quality of experience (QoE) due to increased temperature and shorten battery lifespan.
Therefore, it is important to optimize the energy efficiency of LLM inference.

Since computation and memory access (mainly weights loading) are two major causes of the large energy consumption, there have been extensive works to optimize these two aspects, from both software and hardware perspectives. We have introduced various types of software optimizations in previous sections. For example, model compression methods save energy by reducing the model size and computations; KV cache saves energy by avoiding redundant computations; efficient attention kernels also improve energy efficiency through memory reuse and locality optimizations.

Besides software optimizations, utilizing energy efficient hardware provides new opportunities to improve the agent system's efficiency.
While CPUs and GPUs remain mainstream options to run LLM inference on edge devices, they are designed to support general purpose tasks and don't have dedicated optimization for transformer-based models, especially the generative LLMs.
Researchers have explored to utilize efficient processors that are more suitable to LLM inference workloads, including NPUs~\cite{qualcomm_npu} and TPUs~\cite{reidy2023efficient-tpu}.
However, the limited operator and model support remain challenging in the real-world deployment.
Besides, existing works also designed FPGA-based solutions to boost LLM inference with higher memory bandwidth and energy efficiency ratio (EER) \cite{hong2022dfx-fpga, ye2023accelerating-fpga}.

Yet, the research on energy efficiency of LLM inference is still far from insufficient due to the complexity of hardware deployment and the volatility of energy measurement and analysis. There have be several studies that focus on this topic, such as evaluaing LLMs' inference energy on GPUs \cite{Samsi2023FromWT, Stojkovic2024TowardsGL}, edge devices \cite{laskaridis2024melting} and carbon footprint of LLMs in datacenters \cite{Faiz2023LLMCarbonMT}. Other works tent to present fast energy prediction method for LLM inference, such as IrEne \cite{Cao2021IrEneIE}, which conducted layer-level energy analysis on Transformer-based NLP models and gave an interpretable and extensible energy prediction system. However these prediction models are only for GPU host backends and lack of generalization to other hardware platforms such as mobile phones where \mlas are more likely to be deployed.

\begin{remark}
    \textbf{Remark.} How to improve the efficiency of LLM inference has been extensively studied recently. Despite the remarkable progress, there is still a large gap towards the ubiquitous and affordable deployment of \mlas. The open problems are:

    \begin{enumerate}
        \item Is it possible to further compress or design highly compact models without accuracy degradation, surpassing the scaling law of language models?
        \item If the scaling law is unbreakable, how can we achieve optimal tradeoffs between efficiency and quality via dynamic inference (\eg dynamic collaboration of big model and small model)?
        \item How would the hardware and operating systems evolve to accommodate the efficient deployment of LLMs and \mlas?
    \end{enumerate}
\end{remark}

\subsection{Efficient Customization}

\label{sec:customization-efficiency}

The \mlas may need to serve different users, different tasks, and different scenarios with the same base LLM, which requires efficient customization for each situation.
There are mainly two ways to customize the behaviors of LLMs; one is feeding the LLM with different contextual prompts for in-context learning, and another is tuning the LLM with domain-specific data.
Therefore, the efficiency of customization is primarily determined by the context loading efficiency and LLM fine-tuning efficiency.


\subsubsection{Context Loading Efficiency}

\label{sec:context-loading-efficiency}

Frequent context loading is inevitable during the multi-task serving of \mlas, where each task or each scenario may require a new context for LLM inference.
Nevertheless, the stringent resource constraints inherent to personal devices pose a significant challenge for \mlas to process cumbersome context information fast and efficiently.
There are various ways to make the context loading process more efficient. A straightforward way is to prune some redundant tokens or shorten the context length, which have been discussed in Section~\ref{sec:inference-efficiency}.

Another way to boost context loading is to reduce the bandwidth consumption during context data transmission. In some cases, pruning or discarding some tokens inevitably hurts the LLMs' performance and loading the KV cache necessitates high bandwidth cost. CacheGen \cite{liu2023cachegen} addresses the challenges posed by context loading and it leverages the distinct characteristics of KV features across both tokens and layers thus introduces a novel KV encoder design. This encoder proficiently compresses the KV cache into a compact bitstream, effectively curtailing bandwidth demands while simultaneously reducing processing latency.
Besides, given the fact that different input prompts may have overlapping text segments, \citet{Gim2023PromptCM} proposes Prompt Cache to reuse attention states across prompts. By pre-computing and storing the attention states of frequently occurring text, the framework can efficiently reuse them when these segments appear in new prompts, thus accelerating the inference process.

\subsubsection{Fine-tuning Efficiency}

\label{sec:fine-tuning-efficiency}
It is also desirable to fine-tune a base LLM to better support domain-specific tasks, which poses a significant challenge on computational resources and memory footprint owing to the vast number of parameters in LLMs.
There has been various efforts to tackle these problems, which can be roughly categorized as \textit{parameter-efficient fine-tuning techniques}, \textit{efficient optimizer design} and \textit{training data curation}, which will be elaborated in the following sections.

\textbf{Parameter-efficient fine-tuning (PEFT).} A huge amount of parameters in LLMs make it costly to conduct full-parameter fine-tuning. Lots of efforts on parameter-efficient fine-tuning emerged to reduce LLMs' training overhead. The fundamental concept of PEFT is to freeze the majority of parameters, focusing solely on training a limited set or introducing an adapter with significantly fewer parameters.
A common practice is to introduce some adapters, \ie small neural networks modules, into the existing network structure, including tuning hidden states \cite{houlsby2019parameterefficient,hu2023llmadapters,he2022unified}, adding full layers \cite{houlsby2019parameterefficient} and prepending some prefix vectors into transformer architecture \cite{li2021prefixtuning,liu2022ptuning,zhang2023llamaadapter}. \citet{liu2023gpt} also incorporates trainable vectors at the input layer, the performance of which highly depends on the capabilities of the underlying models. Some of these works fail to avoid extra adapter computation and introduce inference latency. LoRA \cite{hu2022lora} freezes all the model weights and augments each transformer layer with additional rank decomposition matrices, greatly reducing the memory and storage usage during fine-tuning without any additional inference latency.
Another advantage of LoRA is that users can easily switch between different downstream tasks by simply adding or subtracting adapter matrices.
\(\mathtt{(IA)^3}\) \cite{Liu2022FewShotPF} explores element-wise multiplication of the model's activations against learned vectors. It introduces learned vectors which rescale the keys and values in attention mechanisms, and the inner activations in position-wise feed-forward networks. By only training the vectors, \(\mathtt{(IA)^3}\) could maintain the performance with much less computation.

\textbf{Efficient Optimizer Design.} Efficient optimizer design is another group of training/fine-tuning strategies which aims to accelerate the training or reduce the memory overhead during training. Sophia \cite{liu2023sophia}, a lightweight second-order optimizer, addresses the high cost and time required for LLM pre-training by providing a more efficient optimization process compared to commonly used methods like Adam and its variants. On the other hand, the huge number of parameters necessitates storing more activation and optimizer states especially in larger batch size, which places substantial memory demands. LOMO \cite{lv2023parameter} presents a detailed analysis of the memory profile, throughput, and downstream performance of the proposed optimizer compared to other methods, demonstrating significant reductions in memory usage while maintaining training efficiency.
\citet{Zhao2024SecondOrderFW} propose HiZOO, aimed at leveraging the diagonal Hessian to enhance zeroth-order optimizer for fine-tuning LLMs. It avoids the expensive memory cost with one more forward pass per step.

\textbf{Training Data Curation.} Aforementioned approaches primarily focus on the process of training LLMs, while there are also some studies that aim to enhance the LLMs' training performance from a distinct perspective, \ie \textit{the amount and quality of training data}. It has been demonstrated in phi-1 \cite{gunasekar2023textbooks} that training the LLMs with a small amount of high-quality data can lead to significantly reduced training cost and achieve capabilities comparable to large-scale datasets and models.
This challenges the traditional scaling laws in deep learning that emphasize larger datasets and models. Furthermore, phi-1.5 \cite{li2023textbooks} and phi-2 \cite{li2023phi2} extend their focus on many other kinds of tasks such as common sense reasoning and language understanding, achieving comparable performance to models 5x and 25x larger, respectively.
Similarly, TinyGSM \cite{Liu2023TinyGSMA} introduced a synthesized dataset with a few amount (12.3M) of samples on grade school math, which led to remarkable accuracy when tuning small language models with the dataset.

Notably, these methods often assume that the LLMs can fit entirely within the device memory, which isn't a practical assumption for \mlas deployed on personal devices which usually have limited computing power and memory capacity. Fine-tuning LLMs on these devices often requires leverage of hierarchical storage like CPU memory even disk storage. Therefore, when fine-tuning LLMs on personal devices, it's important to carefully consider the resource limitations of the current system.
\begin{remark}
    \textbf{Remark.} While efficient model fine-tuning and in-context learning techniques have been extensively studied, it is yet unclear what is the ideal mechanism for customizing \mlas under different situations.
    Here we highlight two open problems that may be specifically important in the system for \mlas.

    \begin{enumerate}
        \item Similar to the operating system that manages the RAM for the applications, how should the agent system efficiently manage the contexts for different (and potentially parallel) agents, tasks, and users?
        \item Similar to mobile apps that can be efficiently installed, uninstalled and moved between devices, how can a customized (fine-tuned) agent efficiently roll back to the previous versions or transfer to other base models?
    \end{enumerate}
\end{remark}

\subsection{Efficient Memory Manipulation}
\label{sec:memory-retrieval-efficiency}

The \mlas need to frequently retrieve external memory to enable more informed decisions, which can depend on the prevailing mechanism called Retrieval-Augmented Generation(RAG). Considering the diverse forms of external memory data, such as user profiles, interaction history, and local raw files (images, videos, etc.), the common practice is to use embedding models \cite{Word2vec,doc2vec} to represent memory data with a uniform and high-dimensional vector format. The distance between vectors stands for the semantic similarity between the corresponding data. For each given query, the \mlas need to find the most relevant content in external memory storage. The retrieval knowledge then will be injected into \mlas through either prompt concatenation or intermediate layer cross-attention \cite{Zhang2024AcceleratingRL}, with both ways complicating the context of LLM inference. This leads to LLM conducting more efficient computations over long contexts and trying to minimize the memory footprints while undergoing inference, which are similar to improving inference efficiency of LLM as discussed in Section~\ref{sec:inference-efficiency}.

Therefore in this subsection, we mainly focus on the efficient external memory retrieval, which can be considered from two aspects: efficient search and efficient workflow. Efficient search focuses on vector indexing and fast search inside structures like vector libraries (like Faiss \cite{retallm, melz2023enhancing, Zhong2022TrainingLM} and SCaNN \cite{borgeaud2022improving}), vector databases \cite{comprehensiveVDBMs, VDBsurvey, VectorDMs}, or some customized memory structures \cite{Wu2022MemorizingT, modarressi2023ret} where external memory is stored. While efficient workflow targets to further optimize the end-to-end efficiency of retrieval augmented LLM inference.

\subsubsection{Search Efficiency}
\label{sec: Efficient Search}
When comparing the similarity between query vector $q$ and vectors in external memory, a brute-force approach results in a computational complexity of $O(DN)$. However, this approach becomes impractical for scenarios with large vector dimensions ($D$) and dataset sizes ($N$). To alleviate the searching overhead, indexing is commonly employed to expedite query searching by reducing the number of required comparisons.

\textbf{Typical Indexing Algorithms.} This is achieved through partitioning schemes \cite{VDBsurvey} that divide the dataset $S$ into smaller subsets, facilitating selective comparisons and faster search query processing. These partitions are then organized into data structures such as tables, trees, and graphs to enable efficient traversal. Commonly used partitioning methods include randomization (such as RPTree \cite{RPTree1, RPTree2} and E2LSH \cite{E2LSH}), learned partitioning (such as SPANN \cite{spann}), and navigable partitioning (such as NSW \cite{NSW} and HNSW \cite{HNSW}). These partitioning methods can be utilized in combination with different data structures. For example, Vamana \cite{Filtered-DiskANN} is a monotonic search network that comes in graph indexing and uses random initialization.

\textbf{Hardware-aware Index Optimization.}
Since improving the scalability and efficiency of indexing has become a critical concern, research efforts have also focused on hardware-aware approaches to extend external memory capacity while maintaining low latency and high throughput. This is achieved through the utilization of disk-based indexes or the co-design of hardware and algorithms \cite{YaoTianApproximateNN}. For example, DiskANN \cite{Subramanya2019DiskANNFA} addresses cost-effectiveness by employing a hybrid DRAM-SSD approach. It incorporates Vamana graph indexing on SSDs and employs compressed point representation in DRAM. This configuration enables accurate query responses with less than 10ms latency, even when dealing with a billion-point database. DiskANN++ \cite{DiskANN++} further improves efficiency by introducing dynamic entry vertex selection and optimizing SSD layout. This enhancement results in a 1.5x to 2.2x increase in Query Per Second (QPS) while maintaining accuracy on real-world datasets. Moreover, CXL-ANNS \cite{CXL-ANNS} introduces a collaborative software-hardware approach for scalable approximate nearest neighbor search (ANNS). By utilizing Compute Express Link (CXL), CXL-ANNS disentangles DRAM from the host and consolidates essential datasets into its memory pool. FANNS \cite{FANNS} is a vector search framework on FPGAs, featuring automatic co-design of hardware and algorithms based on user-defined recall requirements and hardware constraints. It supports scale-out with a hardware TCP/IP stack and exhibits notable speedups compared to FPGA and CPU baselines.

In terms of the efficiency analysis and optimization of searching itself, some aspects are related to search mechanism design, such as similarity measurement, searching scope, as well as query types, selection, and optimizations. While some aspects, on the other hand, focus on efficient execution of the search process.

\textbf{Search Mechanism Design.}
Multiple similarity criteria can be employed to evaluate vector similarity, including Hamming Distance, Cosine Distance, and Aggregate Scores \cite{Milvus}. However, the selection of scoring mechanisms lacks stringent principles and often relies on empirical rules \cite{VDBsurvey}. Regarding the types of searches, both approximate and exact $k(\geq 1)$ nearest neighbors \cite{YaoTianApproximateNN} search, as well as distance range search, can be utilized to retrieve corresponding vectors. To optimize search latency, rule-based \cite{qdrant, vespa} or estimated-cost-based methods \cite{AnalyticDB-V, Milvus} are often employed to determine the optimal search plan. These rules and cost models are typically configured offline to avoid unnecessary or time-consuming search actions. To further optimize the search process, hybrid operations that combine vector search with metadata filters are gaining popularity. This involves techniques such as pre-filtering \cite{AnalyticDB-V, Milvus, Filtered-DiskANN}, post-filtering, and single-stage filtering \cite{HQANN} to narrow the scope of vector searching. 

\textbf{Search Process Execution.} Several hardware acceleration methods can be taken to improve the efficiency of search executions. For example, to enable parallel query process, Faiss \cite{faiss} uses OpenMP multi-threading, while Milvus \cite{Milvus} further reduces CPU cache misses and uses a novel fine-grained mechanism to best leverage multi-core parallelism. Furthermore, Faiss and Quicker ADC \cite{quickeradc} also support SIMD shuffle instruction to parallelize these table look-ups within a single SIMD processor. GPU is also used for fast query processing \cite{Zhao2020SONGAN, Groh2019GGNNGG, Ootomo2023CAGRAHP}, such as vector databases like Faiss, and Milvus. Many vector database management systems also support distributed clusters to scale to larger datasets or heavier workloads, such as Vald \cite{vald}, Qdrant \cite{qdrant}, etc.

\subsubsection{Workflow Optimization}
\label{sec: Efficient Workflow}
No matter for one-shot or iterative RAG system, traditional workflow is sequential, with inference/retrieval stage idle while conducting retrieving/generation. This feature ignores chances of optimization from the potential of execuation parallelism and retrieval locality of requests. Recent studies are working on pipeline and cache techniques to further improve the efficiency of RAG systems.

\textbf{Pipelining.}
RaLMSpec \cite{Zhang2024AcceleratingRL} is the first work to leverage the advantage of pipeline by enabling a local cache for speculative retrieval. To maintain correctness, a batched verification step is used to guarantee correctness. Besides, cache prefetching, optimal speculation stride scheduler, and asynchronous verification are adopted to further boost the speculation performance. PipeRAG \cite{jiang2024piperag} also uses pipeline, and enhance its performance with two different solutions: flexible retrieval intervals and a performance model informed to dynamically adjust the vector search space depending on the latency expectation of the upcoming token in LLM inferences in the pipeline. PipeRAG utilizes an algorithm-system co-design to avoid increasing end-to-end generation latency while optimizing the search quality.

\textbf{Caching.}
The reason of selecting cache method arises from the temporal and spatial locality of retrieved documents during different requests, which RaLMSpec \cite{Zhang2024AcceleratingRL} has already been utilizing. RAGCache \cite{jin2024ragcache} further uses knowledge tree to organize the intermediate states of the retrieved documents both in the GPU and host memory hierarchy. It also presents a prefix-aware Greedy-Dual-Size-Frequency (PGDSF) replacement policy and a cache-aware request scheduling approach to minimize the cache miss rate. Another work, GRITLM \cite{muennighoff2024generative}, trains LM to handle both generative and embedding tasks by distinguishing between them through instructions. Since the common scenario in RAG is an embedding model used for providing relevant context to the generative model to answer user queries, with GRITLM, the embedding and generative model are equivalent, allowing us to conduct Query Caching or Query-Doc Caching and save computation overhead.



\begin{remark}
    \textbf{Remark.} Managing memory data with external vector storage is not a new requirement for LLM agents.
    While many basic technical challenges have been adequately addressed, we point out two problems that demand specific consideration for \mlas.
    \begin{enumerate}
        \item \mlas may frequently update the memory. Thus, the external memory is expected to facilitate fast updates, maintenance, and re-indexing.
        \item The memory of \mlas may be stored on personal devices with limited storage space, while the memory of the personal agents will accumulate over time. Therefore, it is necessary to effectively compress the memory to avoid fast-growing space and computational cost.
    \end{enumerate}
\end{remark}

\newcommand{\oracle}[1]{\textcolor{blue}{(oracle)#1}}
\newcommand{\checked}[1]{\textcolor{green}{}}
\newcommand{\yile}[1]{\textcolor{orange}{(Yile){#1}}}



\section{Security and Privacy}

\begin{figure}[ht]
  \centering
  \includegraphics[width=15cm]{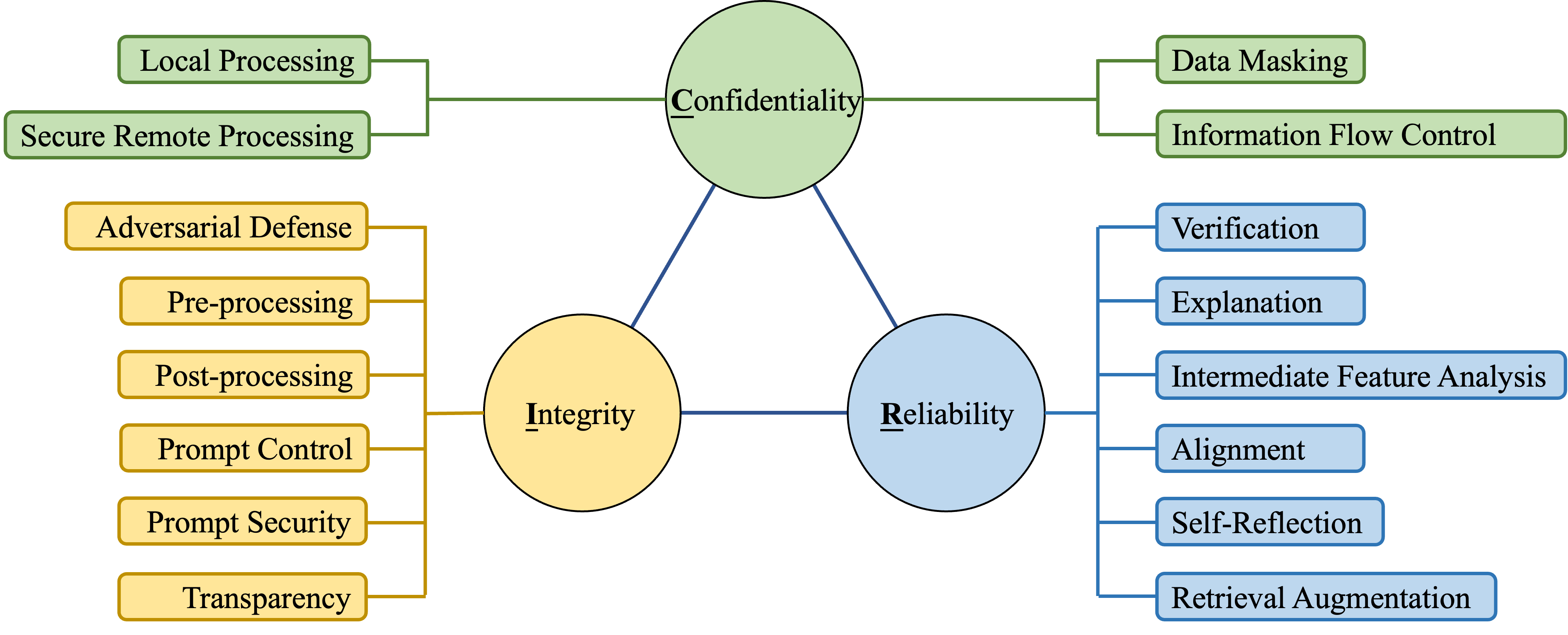}
  \caption{The summary of techniques to address security and privacy issues of \mlas.}
  \label{fig:security}
\end{figure}

The extensive integration of sensitive personal data and safety-critical personal tools sets \mlas apart from regular LLM agents. As a result, ensuring the protection of user data privacy and service security in \mlas becomes a crucial problem.
In the context of \mlas, we focus on three security principles including confidentiality, integrity, and reliability, as shown in Figure~\ref{fig:security}. 
\textbf{Confidentiality} represents the protection of user data privacy, ensuring that unnecessary and unauthorized disclosure of sensitive information does not occur during user interactions with the agents.
\textbf{Integrity} represents the resilience of the agents' decisions, ensuring that the behaviors performed by the agent align with the intended behaviors and have not been deliberately modified or influenced by malicious parties.
\textbf{Reliability} focuses on making the agents' behaviors more dependable and truthful. Unlike integrity, where incorrect answers are a result of intentional external manipulation, reliability addresses the agents' internal mistakes.
\checked{}

\subsection{Confidentiality}

In this subsection, we discuss possible methods for protecting user privacy in \mlas. As mentioned earlier, ensuring user privacy is of utmost importance for the personal agents that have access to a significant amount of user-sensitive data. Unlike traditional LLM-based chatbots where the users explicitly input text, \mlas have the potential to spontaneously initiate queries in places without user awareness, which may contain sensitive information about the user. Meanwhile, the agents may also expose the user information to other agents or services. Consequently, the protection of user privacy becomes even more critical.
There are various methods to enhance the confidentiality, including local data processing, homomorphic encryption, data masking, permission access control, etc.

\subsubsection{Local Processing}
A simple and effective approach to protect user privacy is to perform the computations locally on the users' personal devices. While LLM service providers are currently working towards improving security and building user trust, it is important to acknowledge that transmitting private data to the cloud inherently introduces additional potential risks. Therefore, processing all data locally is considered a more secure method of interacting with LLMs compared to transmitting data to the cloud. However, deploying LLMs locally poses challenges in efficiently processing user requests due to resource constraints on personal devices. This can lead to slow inference speed or even the inability to perform inference due to the limitations of available memory. 
Since the data in \mlas is mainly processed by the LLM, the key to achieve local computation is to run the LLM on users' own devices.
There are various existing lightweight models \cite{touvron2023llama,li2023textbooks} and deployment frameworks \cite{2023bluelm, mlc-llm, lightweightAgentSystem} available for deploying models on edge devices. 
Furthermore, various model compression techniques \cite{sparseQuant,xiao2023smoothquant,frantar2022gptq} are proposed to reduce the model size to further enable the local deployment as discussed in section \ref{sec:memory-efficiency}.

Nevertheless, despite the various efforts of researchers, using a locally-deployed model inevitably faces the challenge of limited model accuracy \cite{kaplan2020scaling}.
Most of the domain experts also suggest to adopt a cloud-edge-collaborated deployment approach to achieve better performance tradeoffs.
Meanwhile, like other software applications, many \mlas would also need to communicate with the cloud to provide online services. It is usually difficult or even impossible to keep the private data completely on local devices.







\subsubsection{Secure Remote Processing}

To invoke cloud-based model inference services while preserving privacy, an ideal solution is homomorphic encryption (HE) \cite{HEBase,gentryFHE}.
In this method, the client employs encryption to encode the user's plaintext request, and the server conducts model inference on the resulting ciphertext. Subsequently, the client receives the inference results in the encrypted format and gets plaintext results after decryption. There have been several studies \cite{cryptonet} that have demonstrated the feasibility of applying HE to Deep Neural Networks, showcasing the potential for integrating HE into models.

When employing HE in \mlas, two challenges arise. 
The first challenge pertains to the limitation that not all operations within the LLMs can be executed using HE. HE atmost supports an unlimited number of additions (equivalent to XOR in a boolean circuit) and multiplications (equivalent to AND in a boolean circuit). However, certain operations in the LLMs, such as max, min, and softmax, cannot be accurately performed using HE.
The second challenge involves the slow inference speed associated with HE, given the large computational complexity of LLMs.

There are several solutions to address these two problems.
The-x~\cite{HETheX} presents a workflow for replacing original non-linear layers with layers that can be computed using HE. In cases where HE cannot perform certain operations, such as the Max operation, the ciphertext will be sent back to the local device. The local device will then perform the operation and send the re-encrypted text back to the cloud. Cheetah~\cite{HECheetah} encompasses a collection of algorithmic and hardware optimizations designed for HE inference on server-side systems. The primary objective of Cheetah is to enhance the computational efficiency of HE, thereby accelerating the speed of HE operations.

However, despite the numerous efforts on accelerating HE-based DNN inference, the current state of homomorphic encryption still falls significantly short of meeting the latency demands of agents \cite{HESurvey}. \checked{}


Beside HE, Multi-Party Communicatio (MPC) \cite{MPC} is an important part of traditional applied cryptography, which refer to communication processes involving multiple parties, where several participants need to communicate in a untrusted environment. 
The challenge of applying MPC in LLM lies in the high computation cost and the significant transition from the mathematical theory of MPC to the actual implementation on LLM. Crypten \cite{crypten} is a framework that includes common MPC methods, supports standard PyTorch tensor operations, and enables GPU computations.

Another way to achieve confidential remote data processing is using the trusted execution environments (TEE) \cite{tramer2018slalom} for model inference. However, TEE may be subject to various attacks \cite{fei2021security_sgx} and may also lead to limited performance.

\subsubsection{Data Masking}

An alternative approach is using data masking to preprocess the information before sending to the cloud. The basic idea is to transform the original inputs into a form that is not privacy-sensitive while preserving the information that has a crucial impact on the inference results.\checked{}

One direct approach of data masking is to transform the plaintext inputs by hiding or replacing sensitive content such as account numbers, addresses, and personal names. These types of information are commonly referred to as Personally Identifiable Information (PII). However, accurately defining PII can be challenging due to its obscure boundaries and diverse forms, making it difficult to consistently identify and remove it from the original content. The National Institute of Standards and Technology (NIST) has provided a guide \cite{PIIguide} that offers recommendations for safeguarding the confidentiality of PII, which could help manage PII more securely. EmojiCrypt\cite{emojicrypt} suggested to use emoji to replace user sensitive information, and then use the modified sentences for generation.

On the other hand, researchers have proposed embedding-based data anonymization approaches where the client encodes the original user request into hidden vectors and sends these vectors to the cloud-based model for subsequent inference. The challenge is how to ensure privacy is protected, how to ensure inference accuracy will not degrade, and how to ensure the inference speed will not decrease too much. There are several solutions.\citet{EmbeddingTrainingTargetModify} propose a metric to assess the extent of privacy leakage in neural representations and develop a defense method by altering training objectives to achieve a tradeoff between privacy and accuracy. \citet{EmbeddingDynamicFusion} protects user privacy by adding dynamic fusion to the intermediate representation. 
TextObfuscator~\cite{embeddingObfuscation} protects user privacy through text obfuscation techniques.
During the encoding process, ``adversarial representation learning'' can be employed by introducing additional constraints to minimize the inclusion of privacy-sensitive information in the encoded vectors~\cite{ALUM}.
Although this method outperforms Homomorphic Encryption in terms of inference performance, it usually does not rigorously protect the data privacy, as the encoded vectors themselves still carry a risk of leaking sensitive information.
Additionally, such methods require an explicit definition of privacy features for the encoder to learn how to remove privacy information during adversarial representation learning.\checked{}

\subsubsection{Information Flow Control}

The aforementioned techniques primarily pertains to the privacy of model input data, while there may also exist the risks of privacy leakage in the model output. This is because the output of the model may not only returns directly to the user but also be sent to other third-party applications, models, users, or intelligent agents. For instance, when an intelligent agent assists a user in making restaurant reservations, it may take the user's basic profile and schedule information and feed them into the restaurant reservation software. 
Similarly, when businesses aim to recommend products to users, they may rely on user preference information retrieved from the output of certain personal agents. This method of obtaining privacy information from the output of LLMs is similar to personal data access interfaces in traditional operating systems, where it is crucial to ensure the control and transparency of privacy data access with permission management systems~\cite{permission}. Transparency necessitates informing users about access information regarding privacy data, including the accessing entity (\textit{who}), content (\textit{what}), time (\textit{when}), intent (\textit{why}), access method (\textit{how}), etc. \citet{llmSystemEval} proposed a way of evaluating privacy leaks in LLM-integrated systems.

One can also directly ask the LLMs to retain private information. However, since LLMs work statistically rather than based on explicit rules, their security cannot be rigorously proven. Therefore, we should not consider LLMs as a part of the Trusted Computing Base (TCB) when dealing with data confidentiality.
Therefore, we may need rule-based permission control to constrain what LLMs can do and what LLMs can access. 
Permission mechanisms allow users to configure whether different entities are permitted to access different types of information. In \mlas, one of the challenges in designing permission mechanisms lies in delineating the types of privacy data, as the content obtained by third-party applications is generated by the model. In traditional systems, researchers have proposed numerous methods for fine-grained privacy content subdivision and permission control, as well as privacy data traceability techniques based on information flow propagation~\cite{traditionalPermission}. However, establishing privacy data traceability for the output generated by LLM agents remains an open issue. \checked{}

\begin{remark}
\textbf{Remark.} Ensuring the confidentiality of user data is crucial for \mlas to build user trusts. However, existing privacy protection techniques are still not sufficient to support agents with higher levels of intelligence. There are following open problems:

\begin{enumerate}
  \item Existing approaches face a common challenge to balance efficiency and effectiveness. For example, how can we enable powerful and efficient local LLMs, how can we scale homomorphic encryption (HE) or trusted execution environment (TEE) to large models, and how can data masking/obfuscation techniques achieve rigorous confidentiality?
  \item As a new software paradigm, it is still unclear what is the systematic privacy protection mechanism for \mlas. Do we still need symbolic rules or permissions for access control? How can they seamlessly integrate with the uninterpretable nature of LLMs?
\end{enumerate}

\end{remark}
\subsection{Integrity}

Integrity refers to the capability of \mlas to ensure that it can output the intended content correctly, even when faced with various types of attacks. As \mlas necessitate interactions with diverse data, applications, and other agents, there is a potential presence of hostile third parties seeking to steal user data and assets or disrupt the system's normal function through unconventional means. Therefore, the system must be able to resist various types of attacks. Traditional attack methods such as modifications to model parameters, theft, and tampering of local data could be defended against using encryption, permissions, hardware isolation, and other measures. However, in addition to defending against traditional attack methods, attention should also be paid to new types of attacks that the LLM agents may encounter: adversarial attacks, backdoor attacks, and prompt injection attacks.\checked{}

\subsubsection{Adversarial Attacks}
Malicious attacks primarily achieve their objectives through the specialized customization of the model's inputs or malicious tampering with the model. A significant category of attacks, known as ``adversarial attacks'', causes model inference errors by customizing or tampering with the model's input data, which was initially discovered in image classification models \cite{szegedy2014intriguing}.
This type of attacks can induce serious classification errors by adding imperceptible noise to images. Subsequently, researchers have extended this attack method to text data, graph data, and beyond \cite{xu2020adversarial_attack_review}. Such attacks also persist in large langage models \cite{kumar2023certifying}, which may also accept input of images \cite{zhao2023evaluating}, text \cite{wei2023jailbroken}, and other modalities of data \cite{schlarmann2023adversarial} from third parties. 
For example, when assisting users in automating tasks, attackers may misguide the agent to delete calendar events and leak private conversation data~\cite{fu2023misusing}, because LLMs often need to input the content of the application's internal information to generate the next interaction decision. In such cases, if the third-party application feed the LLM with maliciously customized content, it could drive the intelligent agent to engage in unsafe interaction.
Traditional defense methods against such attacks in deep learning models usually encompass adversarial defense, abnormal input detection, input preprocessing, output security verification, and more \cite{xu2020adversarial_attack_review}. While these methods theoretically remain applicable to LLM and LLM agents, the large scale of parameters and the characteristics of autoregressive generation may render some computationally expensive methods (such as formalized output security validation and detection of anomalous data based on intermediate layer activations) challenging to implement.
Furthermore, some defense methods may require adjustments in the context of LLM. For instance, training the LLM may incur substantial costs, making it impractical to enhance security through adversarial training. Therefore, exploring how to achieve good effects of adversarial defense through parameter-efficient fine-tuning is worth investigating.
\citet{zhu2023autodan} show that current solutions may be too optimistic: defending against these attacks is possible: adversarial attacks generate unlimited but unreadable gibberish prompts, detectable by perplexity-based filters; manual jailbreak attacks craft readable prompts, but their limited number due to the necessity of human creativity allows for easy blocking. Then they introduce AutoDAN, an interpretable, gradient-based adversarial attack that merges the strengths of both attack types. Guided by the dual goals of jailbreak and readability, AutoDAN optimizes and generates tokens one by one from left to right, resulting in readable prompts that bypass perplexity filters while maintaining high attack success rates, which offers a new way to red-team LLMs and understand jailbreak mechanisms via interpretability.

\subsubsection{Backdoor Attacks}
Another common form of attack is the backdoor attack. Traditional model backdoor attacks are often achieved through data poisoning \cite{gu2019badnets}, \ie inserting maliciously modified samples into the model's training data, enabling the model to learn deliberate hidden decision logic, such as ``when seeing an apple pattern, the model outputs an incorrect classification''. For LLMs, data poisoning may be more challenging due to the huge amount and strict unified management of training data, but another type of backdoor attack methods \cite{yuan2023patchbackdoor} is still valid, which implants insecure logic into the model by modifying the model input during the test time.
\citet{kandpal2023backdoor} elicits targeted misclassification when the language models are prompted to perform a particular target task.
ProAttack~\cite{zhao2023prompt} directly utilizes prompts as triggers to inject backdoors into LLMs, which is the first attempt to explore clean-label textual backdoor attacks based on the prompt.
PoisonPrompt~\cite{yao2023poisonprompt} is a bi-level optimization-based prompt backdoor attack on soft and hard prompt-based LLMs. 
Since LLMs often use several fixed prompts in certain scenarios, this form of attack, achieved by modifying the prompts, essentially fine-tunes the model's parameters and thus alters its decision logic.
\citet{han2024effectiveness} distill benign knowledge from poisoned pre-trained encoders and transfer it to a new encoder, resulting in a clean pre-trained encoder, which may hurt the LLMs' performance.
\citet{sun2023defending} proposed that testing the backward probability of generating sources given targets yields effective defense performance against different types of attacks.
Indeed, when attackers mimic normal behavior, this defense method may become ineffective. Therefore, there isn't a robust solution for backdoor defense in agent systems yet~\cite{Abdelnabi2023llmdefense}. 
This highlights the request of developing effective defenses against sophisticated attacks that mimic legitimate behavior.

\subsubsection{Prompt Injection Attacks}

In the era of LLM, there emerges a new and particularly crucial security risk, namely prompt injection attacks \cite{perez2022ignore,liu2023prompt,shayegani2023jailbreak,chao2023jailbreaking}. In this form of attack, the model itself incorporates certain security safeguards through alignment and prompts. Nevertheless, third-party model users can bypass these preset security safeguards by using subtle or special diction in the prompts. For instance, an intelligent personal assistant may be preset not to execute certain sensitive operations, such as modifying a user's account password~\cite{Nicholas2021leak}, but through prompt injection (\eg requesting the LLM to ``disregard the previously set limitations'' or ``assume operation in an authorized secure mod''), it could induce the model to violate regulations and perform these sensitive operations.

For such prompt-based attack methods, there are currently no perfect defense mechanisms. 
SmoothLLM~\cite{robey2023smoothllm} is the first general-purpose defense method for prompt injection, and it randomly perturbs multiple copies of a given input prompt and then aggregates the corresponding predictions to detect adversarial inputs.
However, its defensive effectiveness is highly dependent on the model's robustness, since there was only about 1\% reduction in the attack success rate for some models.
An essential way to mitigate this issue is to ensure the transparency and security of the LLM's prompts. For example, a \mla could rigidly control the template and specifications of prompts, requiring all requests to comply with the preset template and specifications. Additionally, post-processing of the input content from third-party applications (summarization, translation, restatement, etc.) or prompt encapsulation (such as adding explicit text before and after to indicate their origin from a third party) can help the model clearly distinguish them from the system's inherent prompts.

\begin{remark}
\textbf{Remark.}
Ensuring the integrity of the decision process is crucial for \mlas. 
The threats to integrity are very diverse and continuously evolving, while the development of defensive techniques are lying behind. Here we highlight two important open problems that apply to all types of attacks.
\begin{enumerate}
\item 
How can the agents know if their input or decision process has been tampered with by third parties? 
This requires the agents to have a sense of what are normal input and behaviors, and have the abilities to recognize the anomalies.
\item 
Since directly avoiding the attacks may be challenging, it would be more practical to consider user verification mechanisms, \ie asking the user to verify when the agents are uncertain. How to design a secure and user-friendly verification mechanism is challenging.
\end{enumerate}
\end{remark}
\subsection{Reliability}

In \mlas, the LLMs determine numerous critical actions, including some sensitive operations such as modifying and deleting user information, purchasing services, and sending messages. Therefore, ensuring the reliability of the agent's decision-making process is crucial. We discuss the reliability of LLMs from three perspectives, including the \textbf{problems} (\ie \textit{where does reliability issues of LLMs manifest from?}), \textbf{improvement} (\ie \textit{how can we make the LLMs' response more reliable?}), and \textbf{inspection} (\ie \textit{how can we deal with the LLM's potentially unreliable output?}).\checked{}

\subsubsection{Problems}

\textbf{Hallucination.} 
LLMs may produce incorrect answers, which can lead to severe consequences. 
In comparison to LLM-based chatbots that directly interact with users via text, \mlas minimize user disruptions by avoiding frequent result verifications, hence amplifying the severity of producing incorrect answers. Researchers have uncovered cases where LLMs generate text that is coherent and fluent but ultimately erroneous. This phenomenon, known as \textit{hallucination} in natural language processing tasks, poses a challenge to personal agents as well. \citet{HallucinationSurvey} delves deeply into the various manifestations of hallucinations in natural language processing tasks. \citet{rawte2023survey} further discusses the hallucinations in multimodal foundation models, providing valuable references for interested readers.\checked{}

\textbf{Unrecognized Operation.} Unlike the hallucination problem that focuses on the ``wrong answer'' produced by LLMs, there are many cases where the responses from these models are ``not even wrong''. For instance, consider the scenario where the LLM is instructed to initiate a phone call by using the format ``\texttt{CALL XXXXXX}''. In response, the LLM may generate a reply ``\texttt{I will make a call to XXXX}'', which accurately conveys the intended meaning but deviates from the specified format, rendering it unexecutable. As we know, the essence of LLMs is language modeling, and the outputs of language models are typically in the form of language. Compared to other LLMs that interact directly with humans, \mlas is required to execute actions. As a result, they have significantly higher requirements for the format and executability of their outputs~\cite{nair2023dera}.
\checked{}

\textbf{Sequential Reliability.}
LLMs are initially pre-trained on sequential data (\ie corpus) and training objectives (\ie left-to-right language modeling task). However, problems in the real world may not be fully addressed sequentially. Achieving sequential reliability poses several challenges, including context preservation, coherence maintenance, \textit{etc}. To better maintain a coherent and meaningful conversation with users and \mlas, we need to elicit the LLMs' ability to think from a global perspective, not solely relying on the previously generated tokens or contexts. On enhancing the ability of thinking and reasoning of LLMs, \citet{tot} propose Tree-of-Thought to generate and conclude over multiple different reasoning paths, \citet{zhang2023cumulative} propose Cumulative Reasoning in a cumulative and iterative manner to solve complex tasks. 
There is also potential for designing the overall plan for solving the task~\cite{lu2023chameleon} or drawing insights from the previous work~\cite{an2023learning,zhu2023large}. 


\subsubsection{Improvement}

The improvement approaches aim to improve the quality of LLM output, thereby enhancing the reliability of LLM-based agents.


\textbf{Alignment.} As LLMs grow in size and complexity, concerns have arisen regarding their potential to generate biased, harmful, or inappropriate content. Alignment methods seek to mitigate these risks and ensure that the behavior of LLMs aligns with ethical and societal norms. One common alignment method is the use of pre-training and fine-tuning~\cite{gururangan-etal-2020-dont,liu2023pre,weifinetuned}. LLMs are pre-trained on vast amounts of text data to learn language patterns and representations. During the fine-tuning phase, the models are further trained on more specific and carefully curated datasets, including human-generated examples and demonstrations. 
This process helps align the models with the desired behaviors by incorporating human values and intentions into their training. Another alignment method is reward modeling, which involves defining and optimizing a reward function that reflects the desired outcomes or behaviors. By providing explicit rewards or penalties for specific actions, LLMs can be trained to generate output that align with those predefined objectives. Reinforcement learning techniques (\eg RLHF~\cite{ouyang2022training}, RLAIF~\cite{lee2023rlaif}, C-RLFT~\cite{wang2023openchat}) can be employed to optimize the model behavior based on these reward signals. oversight and intervention are critical alignment methods. Human reviewers or moderators play a crucial role in reviewing and filtering the outputs of LLMs for potential biases, harmful content, or inappropriate behavior. Their feedback and interventions are used to iteratively improve the model's performance and align it with desired standards. 

\textbf{Self-Reflection.} 
It has been shown that language models can provide probabilities of providing correct answers \cite{languagemodelsknowwhattheyknow}. Inspired by the autonomous operation of LLMs, researchers have suggested leveraging the model's self-reflection to mitigate the problem of incorrect content generation.
\citet{huang2022large} and \citet{madaan2023self} show that LLMs are capable of self-improving with unlabeled data, \citet{shinn2023reflexion} propose Reflexion to let LLMs update through its linguistic feedback. \citet{chen2023teaching} propose Self-Debug to iteratively improve the responses on several code generation tasks. SelfCheckGPT~\cite{selfcheckgpt} allows large models to provide answers to the same input question multiple times and checks the consistency between these responses.
If there are contradictions among the answers, there is a higher probability that the model has generated unreliable content. 
\citet{debateGPT} attempts to improve the reliability of model outputs by enabling multiple large model agents to engage in mutual discussion and verification. There are various ways to combine models, similar to the diverse collaboration methods in the human world. However, just as more employees require increased expenses, having more models entails greater computational power requirements. The above works demonstrate a trend in which LLMs are evolving from mere textual generators to intelligent agents, transitioning from primitive comprehension-based reasoning to reflective reasoning with iterative updates.



\textbf{Retrieval Augmentation.}
LLMs show strong performance across various tasks, however, the parametric knowledge stored in the models could still be incomplete and difficult to update efficiently. Alternatively, retrieval-augmented methods~\cite{borgeaud2022improving,lewis2020retrieval,realm} provide a semi-parametric way to offer complementary nonparametric information, allowing LLMs to draw on retrieved real-world knowledge when generating content, such as Wikipedia, documents, or knowledge graphs \cite{knowledgeGraphSurvey}. This approach offers the advantage of not requiring model modification, facilitates real-time information updates, and allows the traceability of generated results to the original data, thereby enhancing the interpretability of the generated information. Retrieval augmentation has been shown to be effective for traditional pre-trained models such as BERT~\cite{kenton2019bert}. However, for LLMs that already have strong reasoning ability, augmenting the context could also have a negative impact due to irrelevant or noisy information~\cite{shi2023large}. To tackle these issues, \citet{guozhicheng} propose a prompt-guided retrieval method for non-knowledge-intensive tasks, enhancing the relevance of retrieved passages for more general queries. \citet{robustRetrieval} propose Chain-of-Note to improve the robustness when dealing with noisy and irrelevant documents. \citet{asai2023self} propose Self-RAG to enhance factuality through self-reflection. \citet{WangYiLe} propose SKR, a self-knowledge-guided retrieval method to balance external knowledge with internal knowledge. \citet{retrieve3} propose FLICO to filter the context in advance and improve the fine-grained relevance of retrieved segments.
The CRITIC~\cite{critic} framework utilizes LLMs to verify and iteratively self-correct their output through interaction with external tools, such as a calculator, a Python interpreter, and Wikipedia. \citet{zhang2024raft} propose RAFT, a retrieval augmented fine-tuning for improving domain specific question answering. However, these approaches still rely on high-performance texts retriever and  have limited assistance for user requests for which matching content cannot be easily found in external knowledge bases.\checked{}

\subsubsection{Inspection}

The inspection-based approaches, on the other hand, do not interfere the LLM generation process. Instead, it focuses on how to enhance or understand the reliability of agents based on the already generated results.

\textbf{Verification.} Given that the issue of unreliable content generation by LLMs cannot be entirely avoided when deploying such systems for actual use, it remains necessary to establish rule-based security verification mechanisms. 
Regarding the aforementioned unrecognized operation, ``Constrained Generation'' refers to the process of generating formatted and constrained output, which can be employed to tackle this issue. 
\citet{ConstrainedGenerationSampling} employs Langevin Dynamics simulation for non-autoregressive text generation as a solution to this problem. On the other hand, \citet{ConstrainedGenerationModification} introduces a method that suggests a candidate modification at each iteration and verifies if the modified sentence satisfies the given constraints to generate constrained sentences. \citet{li-etal-2023-making} and \citet{weng-etal-2023-large} propose self-verification to help the reasoning process of large language models. Responsible Task Automation~\cite{responsible_task_automation} is a system that can predict the feasibility of commands, confirm the completeness of executors, and enhance the security of large language models.
However, further research is needed to improve the accuracy and recall rates in identifying sensitive operations and to mitigate the decision burden on users. \checked{}

\textbf{Explanation.} While it is mentioned earlier that intelligent personal assistants should minimize user interruptions, incorporating user opinions or human assistance can be valuable, particularly when making significant decisions. In case an intelligent personal assistant makes a mistake, having interpretable logic can also be helpful in the subsequent debugging process. There are several surveys \cite{xai1,xai2,xai3} discussing about explainable language model. Traditionally, rationale-based methods~\cite{carton-etal-2022-learn,gurrapu2023rationalization} can be used to explain the model output by explicitly training on human-annotated data. As for LLMs, chain-of-thought reasoning~\cite{cot} approaches can also help the model generate textual explanations. To make the reasoning process more robust and reliable, recent studies further enhance chain-of-thought reasoning with majority voting~\cite{wang2022self} and iterative bootstrapping~\cite{sun2023enhancing} mechanisms. It is evident that researchers place a significant emphasis on interpretability, as it not only contributes to reliability but also represents an intriguing research direction.

\textbf{Intermediate Feature Analysis.}
Beyond the last-layer representation, some work involves analyzing the intermediate states in the model's inference process to judge the generation of false information. \citet{overthinking} discover that the behavior of a model may significantly diverge at certain layers, highlighting the importance of analyzing the intermediate computations of the model.
\citet{ITI} find that the model activation of intermediate layers can reveal some directions of ``truthfulness'', showing that the LLMs may already capture knowledge though not generated, they further propose shifting the model activation during inference and improving the responses of LLMs. \citet{TheHallucinations} propose a method to leverage mutual information and alleviate hallucination by assessing the confidence level of the next token, where the underlying reason is that the neural activation pattern in LLMs during the generation of hallucinatory content differs from normal outputs. These studies highlight the drawbacks of solely depending on the final-layer representation for language modeling, revealing the potential benefits of harnessing hierarchical information across different layers of the model. \checked{}

\begin{remark}
\textbf{Remark.}
The reliability of LLM generation has received considerable amount of attention, especially around the hallucination problem. However, avoiding the unreliable behaviors is still difficult, if not impossible. The open problems include:
\begin{enumerate}
\item 
How can we evaluate the reliability of LLM and LLM agents? Existing methods rely on either black-box LLMs such as GPT-4 or costly human annotations. Authoritative benchmarks and methods are desired for evaluating and improving the reliability.
\item 
Similar to the confidentiality problem, incorporating rigorous symbolic rules in the decision process of \mlas would be a practical solution for reliability. However, complying with the rules while retaining powerful capabilities of LLM agents is challenging.
\item 
The lack of transparency and interpretability of DNNs has been a long-standing problem, which is even more critical for all security \& privacy aspects of \mlas.
How to interpret and explain the internal mechanisms of LLMs is a direction that worth continuous investigation.
\end{enumerate}
\end{remark}


\section{Conclusion and Outlook}

The emergence of large language models presents new opportunities for the development of intelligent personal assistants, offering the potential to revolutionize the way of human-computer interaction. In this paper, we focus on \mlas, systematically discussing several key opportunities and challenges based on domain expert feedback and extensive literature review.

Currently, research on \mlas is in the early stages. Task execution capabilities are still relatively inadequate, and the range of supported functionalities is rather narrow, leaving significant room for improvement.
Moreover, ensuring the efficiency, reliability and usability of such personal agents requries to address numerous critical performance and security issues. 
There exists an inherent tension between the need of large-scale parameters in LLM to achieve better service quality and the constraints of resource, privacy and security in personal agents.

Going forward, except for addressing the respective challenges in each specific direction, a joint effort is needed to establish the whole software/hardware stack and ecosystem for \mlas. Researchers and engineers also need to carefully consider the responsibility of such technology to guarantee the benign and assistive nature of \mlas.




\section*{Acknowledgment}
\label{sec:ack}

This work is supported by the National Natural Science Foundation of China (NSFC, Grant No.62272261) and collaborative research projects with AsiaInfo Technologies (China) Inc. and Xiaomi Inc.
We sincerely thank the valuable feedback from many domain experts including 
Xiaobo Peng (Autohome), Ligeng Chen (Honor Device), Miao Wei, Pengpeng He (Huawei), Hansheng Hong, Wenjun Chen, Zhiyao Yang (Oppo), Xuesheng Qi (vivo), Liang Tao, Lishun Sun, Shuang Dong (Xiaomi), and the anonymous others.
Among the co-authors, Jiacheng Liu, Wenxing Xu, and Rui Kong were interns at Institute for AI Industry Research (AIR), Tsinghua University when writing this paper.

\bibliographystyle{unsrtnat}  
\bibliography{references}

\end{document}